\documentclass{llncs}

\usepackage{latexsym}
\usepackage{amsxtra} 
\usepackage{amssymb}
\usepackage{amsmath}
\usepackage{todonotes}
\usepackage{enumerate}

\usepackage{color}
\usepackage{pslatex}

\usepackage{epsfig}

\usepackage{pgf}
\usepackage{tikz}
\usetikzlibrary{arrows,shapes,automata,positioning,calc}

\renewcommand{\vec}[1]{{\bf {#1}}}
\newcommand{\seq}[1]{\overrightarrow{\vec{#1}}}

\newcommand{\len}[1]{{|{#1}|}}
\newcommand{\card}[1]{{|\!|{#1}|\!|}}

\newcommand{\lang}[1]{{\mathcal L}({#1})}

\newcommand{\arrow}[2]{\xrightarrow[{\scriptstyle #2}]{{\scriptstyle #1}}}

\newcommand{\nat}{{\bf \mathbb{N}}}

\newif\ifLongVersion\LongVersiontrue

\def\mkstate<#1,#2>{\langle #1,#2 \rangle}

\newcommand{\dir}{\mathcal{D}}

\begin{document}

\title{The Tree Width of Separation Logic with Recursive Definitions}

\author{Radu Iosif\inst{1} \and Adam Rogalewicz\inst{2} \and Jiri
  Simacek\inst{2}} 
\institute{\textsc{Verimag/CNRS}, Grenoble, France
  \and 
  \textsc{FIT}, Brno University of Technology, IT4Innovations Centre of Excellence, Czech Republic}

\maketitle

\begin{abstract}
Separation Logic is a widely used formalism for describing dynamically
allocated linked data structures, such as lists, trees, etc. The
decidability status of various fragments of the logic constitutes a
long standing open problem. Current results report on techniques to
decide satisfiability and validity of entailments for Separation
Logic(s) over lists (possibly with data). In this paper we establish a
more general decidability result. We prove that any Separation Logic
formula using rather general recursively defined predicates is
decidable for satisfiability, and moreover, entailments between such
formulae are decidable for validity. These predicates are general
enough to define (doubly-) linked lists, trees, and structures more
general than trees, such as trees whose leaves are chained in a
list. The decidability proofs are by reduction to decidability of
Monadic Second Order Logic on graphs with bounded tree width.
\end{abstract}

\section{Introduction}
Separation Logic (SL) \cite{reynolds02} is a general framework for
describing dynamically allocated mutable data structures generated by
programs that use pointers and low-level memory allocation
primitives. The logics in this framework are used by an important
number of academic (\textsc{Space
Invader} \cite{spaceinvader}, \textsc{Sleek} \cite{sleek} and
\textsc{Predator} \cite{predator}), as well as industrial-scale
(\textsc{Infer} \cite{infer}) tools for program verification and
certification. These logics are used both externally, as property
specification languages, or internally, as e.g., abstract domains for
computing invariants, or for proving verification conditions. The main
advantage of using SL when dealing with heap manipulating programs, is
the ability to provide compositional proofs, based on the principle of
{\em local reasoning} i.e., analyzing different sections (e.g.,
functions, threads, etc.) of the program, that work on disjoint parts
of the global heap, and combining the analysis results a-posteriori.

\begin{figure}[h]
\vspace{-1mm}
\begin{tabular}{ccc}
\begin{minipage}{70mm}
\[\begin{array}{rcl}
list(hd,tl) & ::= & emp \wedge hd=tl \\
& | & \exists x.~hd\mapsto x * list(x,tl) \\
dll(hd,p,tl) & ::= & emp \wedge hd=tl \\
& | & \exists x.~hd\mapsto (x,p) * dll(x,hd,tl) \\
tree(root) & ::= & emp \wedge root=nil \\
& | & \exists l,r.~root\mapsto (l,r) * tree(l) * tree(r) \\
tll(x,ll,lr) & ::= & x \mapsto (nil,nil,lr) \wedge x = ll \\ 
& | & \exists l,r,z.~ x \mapsto (l,r,nil) * tll(l,ll,z)  \\
& & * tll(r,z,lr)
\end{array}\]
\end{minipage} & \ &
\begin{minipage}{50mm}
\epsfig{file=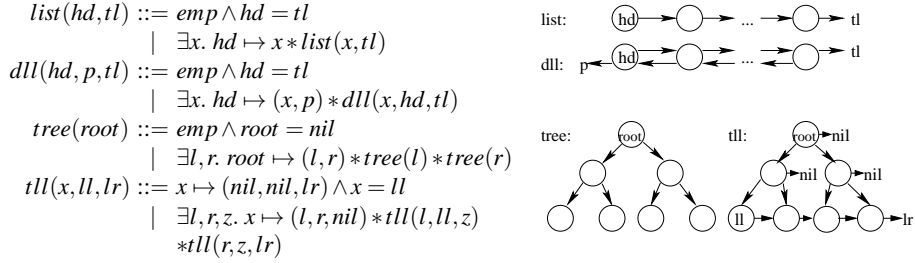,width=50mm}
\end{minipage}

\end{tabular}
\caption{Examples of recursive data structures definable in SLRD.}
\label{fig:definitions}
\vspace{-4mm}
\end{figure}

The basic language of SL consists of two kinds of atomic propositions
describing either (i) the empty heap, or (ii) a heap consisting of an
allocated cell, connected via a separating conjunction
primitive. Hence a basic SL formula can describe only a heap whose
size is bounded by the size of the formula. The ability of describing
unbounded data structures is provided by the use of {\em recursive
definitions}. Figure \ref{fig:definitions} gives several common
examples of recursive data structures definable in this framework.

The main difficulty that arises when using Separation Logic with
Recursive Definitions (SLRD) to reason automatically about programs is
that the logic, due to its expressiveness, does not have very nice
decidability properties. Most dialects used in practice restrict the
language (e.g., no quantifier alternation, the negation is used in a
very restricted ways, etc.) and the class of models over which the
logic is interpreted (typically singly-linked lists, and slight
variations thereof). In the same way, we apply several natural
restrictions on the syntax of the recursive definitions, and define
the fragment $\mbox{SLRD}_{btw}$, which guarantees that all models of
a formula in the fragment have {\em bounded tree width}. Indeed, this
ensures that the satisfiability and entailment problems in this
fragment are decidable {\em without any restrictions on the type of
the recursive data structures considered}.

In general, the techniques used in proving decidability of Separation
Logic are either proof-based (\cite{sleek,berdine-calcagno-ohearn04}),
or model-based
(\cite{bozga-iosif-perarnau10,cook-haase-ouaknine-parkinson-worell11}). It
is well-known that automata theory, through various automata-logics
connections, provides a unifying framework for proving decidability of
various logics, such as (W)SkS, Presburger Arithmetic or MSO over
certain classes of graphs. In this paper we propose an
automata-theoretic approach consisting of two ingredients. First,
$\mbox{SLRD}_{btw}$ formulae are translated into equivalent Monadic
Second Order (MSO) formulae over graphs. Second, we show that the
models of $\mbox{SLRD}_{btw}$ formulae have the {\em bounded tree
width} property, which provides a decidability result by reduction to
the satisfiability problem for MSO interpreted over graphs of bounded
tree width \cite{seese91}, and ultimately, to the emptiness problem of
tree automata.


\subsubsection{Related Work} 
The literature on defining decidable logics for describing mutable
data structures is rather extensive. Initially, first-order logic with
transitive closure of one function symbol was introduced in
\cite{immerman-rabinovich-reps-sagiv-yorsh04} with a follow-up logic
of reachability on complex data structures, in
\cite{yorsh-rabinovich-sagiv-meyer-bouajjani06}. The decision
procedures for these logics are based on reductions to the
decidability of MSO over finite trees. Along the same lines, the logic
\textsc{Pale} \cite{pale2001} goes beyond trees, in defining trees
with edges described by regular routing expressions, whose
decidability is still a consequence of the decidability of MSO over
trees.  More recently, the \textsc{Csl} logic
\cite{bouajjani-dragoi-enea-sighireanu09} uses first-order logic with
reachability (along multiple selectors) in combination with arithmetic
theories to reason about shape, path lengths and data within heap
structures. Their decidability proof is based on a small model
property, and the algorithm is enumerative. In the same spirit, the
\textsc{Strand} logic \cite{madhusudan-parlato-qiu11} combines MSO
over graphs, with quantified data theories, and provides decidable
fragments using a reduction to MSO over graphs of bounded tree
width. 

On what concerns SLRD \cite{reynolds02}, the first (proof-theoretic)
decidability result on a restricted fragment defining only
singly-linked lists was reported in \cite{berdine-calcagno-ohearn04},
which describe a coNP algorithm. The full basic SL without recursive
definitions, but with the magic wand operator was found to be
undecidable when interpreted {\em in any memory model} 
\cite{Brotherston:2010}. Recently, the entailment problem for
SLRD over lists has been reduced to graph homomorphism in
\cite{cook-haase-ouaknine-parkinson-worell11}, and can be solved in
PTIME. This method has been extended to reason nested and overlaid
lists in \cite{enea13}. The logic $\mbox{SLRD}_{btw}$, presented in
this paper is, to the best of our knowledge, the first decidable SL
that can define structures more general than lists and trees, such as
e.g. trees with parent pointers and linked leaves. 


\section{Preliminaries}


For a finite set $S$, we denote by $\card{S}$ its cardinality. We
sometimes denote sets and sequences of variables as $\vec{x}$, the
distinction being clear from the context. If $\vec{x}$ denotes a
sequence, $(\vec{x})_i$ denotes its $i$-th element. For a partial
function $f : A \rightharpoonup B$, and $\bot\notin B$, we denote
$f(x)=\bot$ the fact that $f$ is undefined at some point $x \in A$. By
$f[a \leftarrow b]$ we denote the function $\lambda x ~.~ \mbox{if}~
x=a ~\mbox{then}~ b ~\mbox{else}~ f(x)$. The domain of $f$ is denoted
$dom(f) = \{x \in A \mid f(x) \neq \bot\}$, and the image of $f$ is
denoted as $img(f) = \{ y\in B \mid \exists x\in A ~.~ f(x)=y\}$. By
$f : A \rightharpoonup_{fin} B$ we denote any partial function whose
domain is finite. Given two partial functions $f,g$ defined on
disjoint domains, we denote by $f \oplus g$ their union.

\subsubsection{Stores, Heaps and States.}
We consider $PVar = \{u,v,w,\ldots\}$ to be a countable infinite set
of {\em pointer variables} and $Loc = \{l,m,n,\ldots\}$ to be a
countable infinite set of \emph{memory locations}. Let $nil \in PVar$
be a designated variable, $null \in Loc$ be a designated location, and
$Sel = \{1,\ldots,\mathcal{S}\}$, for some given $\mathcal{S} > 0$, be
a finite set of natural numbers, called {\em selectors} in the
following.

\begin{definition}\label{state}
A \emph{state} is a pair $\langle s, h \rangle$ where $s : PVar
\rightharpoonup Loc$ is a~ partial function mapping pointer variables
into locations such that $s(nil)=null$, and $h : Loc
\rightharpoonup_{fin} Sel \rightharpoonup_{fin} Loc$ is a finite
partial function such that (i) $null \not\in dom(h)$ and (ii) for all
$\ell\in dom(h)$ there exist $k \in Sel$ such that $(h(\ell))(k) \neq \bot$. 
\end{definition}
Given a state $S=\langle s, h \rangle$, $s$ is called the \emph{store}
and $h$ the \emph{heap}. For any $k \in Sel$, we write $h_k(\ell)$
instead of $(h(\ell))(k)$, and $\ell \arrow{k}{} \ell'$ for $h_k(\ell) =
\ell'$. We sometimes call a triple $\ell \arrow{k}{} \ell'$ an {\em
  edge}, and $k$ is called a {\em selector}. Let $Img(h) =
\bigcup_{\ell \in Loc} img(h(\ell))$ be the set of locations which are
destinations of some selector edge in $h$. A location $\ell \in Loc$
is said to be \emph{allocated} in $\langle s, h \rangle$ if $\ell \in
dom(h)$ (i.e. it is the source of an edge), and \emph{dangling} in
$\langle s, h \rangle$ if $\ell \in [img(s) \cup Img(h)] \setminus
dom(h)$, i.e., it is either referenced by a store variable, or
reachable from an allocated location in the heap, but it is not
allocated in the heap itself. The set $loc(S) = img(s) \cup dom(h)
\cup Img(h)$ is the set of all locations either allocated or
referenced in a state $S = \langle s, h \rangle$.

\subsubsection{Trees.}
Let $\Sigma$ be a finite label alphabet, and $\nat^*$ be the set of
sequences of natural numbers. Let $\epsilon \in \nat^*$ denote the
empty sequence, and $p.q$ denote the concatenation of two sequences
$p,q\in \nat^*$.  A \emph{tree} $t$ over $\Sigma$ is a~ finite partial
function $t : \nat^* \rightharpoonup_{fin} \Sigma$, such that $dom(t)$
is a finite prefix-closed subset of $\nat^*$, and for each $p \in
dom(t)$ and $i \in \nat$, we have: $t(p.i)\neq\bot \Rightarrow \forall
0 \leq j < i ~.~ t(p.j) \neq \bot$. Given two positions $p, q \in
dom(t)$, we say that $q$ is the $i$-th successor (child) of $p$ if $q
= p.i$, for $i\in\nat$. Also $q$ is a successor of $p$, or
equivalently, $p$ is the parent of $q$, denoted $p= parent(q)$ if $q =
p.i$, for some $i \in \nat$.

We will sometimes denote by $\mathcal{D}(t) = \{-1,0,\ldots,N\}$ the
{\em direction alphabet} of $t$, where $N = \max\{i \in \nat ~|~ p.i
\in dom(t)\}$. The concatenation of positions is defined over
$\mathcal{D}(t)$ with the convention that $p.(-1) = q$ if and only if
$p = q.i$ for some $i \in \nat$. We denote $\mathcal{D}_+(t) =
\mathcal{D}(t) \setminus \{-1\}$. A {\em path} in $t$, from $p_1$ to
$p_k$, is a sequence $p_1,p_2,\dots,p_k\in dom(t)$ of pairwise
distinct positions, such that either $p_i=parent(p_{i+1})$ or
$p_{i+1}=parent(p_i)$, for all $1 \leq i < k$. Notice that a path in the
tree can also link sibling nodes, not just ancestors to their
descendants, or viceversa. However, a path may not visit the same tree
position twice.

\subsubsection{Tree Width.}
A state (Def. \ref{state}) can be seen as a directed graph, whose
nodes are locations, and whose edges are defined by the selector
relation. Some nodes are labeled by program variables ($PVar$) and all
edges are labeled by selectors $(Sel)$. The notion of tree width
is then easily adapted from generic labeled graphs to
states. Intuitively, the tree width of a state (graph) measures the
similarity of the state to a tree. 

\begin{figure}[htb]
\begin{center}
\epsfig{file=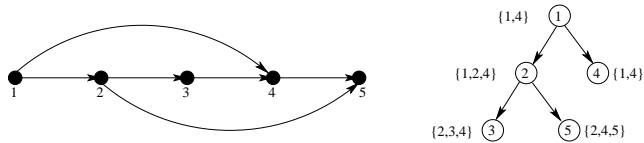}
\end{center}
\caption{A graph and a possible tree decomposition of width 2}
\label{fig:sll-grid}
\end{figure}

\begin{definition}\label{tw}
  Let $S = \langle s, h \rangle$ be a state. A {\em tree
    decomposition} of $S$ is a tree $t : \nat^* \rightharpoonup_{fin}
  2^{loc(S)}$, labeled with sets of locations from $loc(S)$, with the
  following properties:
  \begin{enumerate}
  \item $loc(S) = \bigcup_{p \in dom(t)} t(p)$, the tree covers the
    locations of $S$
  \item for each edge $l_1 \arrow{s}{} l_2$ in $S$, there exists $p
    \in dom(t)$ such that $l_1,l_2 \in t(p)$
  \item for each $p,q,r \in dom(t)$, if $q$ is on a path from $p$ to
    $r$ in $t$, then $t(p) \cap t(r) \subseteq t(q)$
  \end{enumerate}
  The width of the decomposition is $w(t) = \max_{p \in dom(t)} \{
  \card{t(p)} - 1 \}$. The tree width of $S$ is $tw(S) = \min \{w(t)
  ~|~ t ~\mbox{is a tree decomposition of $S$}\}$.
\end{definition}
A set of states is said to have {\em bounded tree width} if there
exists a constant $k \geq 0$ such that $tw(S) \leq k$, for any state
$S$ in the set. Figure \ref{fig:sll-grid} gives an example of a graph (left)
and a possible tree decomposition (right). 

\subsection{Syntax and Semantics of Monadic Second Order Logic}\label{sec:MSO}

Monadic second-order logic (MSO) on states is a straightforward
adaptation of MSO on labeled graphs \cite{madhusudan-parlato11}. As
usual, we denote first-order variables, ranging over locations, by
$x,y,\dots$ , and second-order variables, ranging over sets of
locations, by $X,Y,\dots$. The set of logical MSO variables is denoted
by $LVar_{mso}$, where $PVar \cap LVar_{mso} = \emptyset$. 

We emphasize here the distinction between the logical variables
$LVar_{mso}$ and the pointer variables $PVar$: the former may occur
within the scope of first and second order quantifiers, whereas the
latter play the role of symbolic constants (function symbols of zero
arity). For the rest of this paper, a logical variable is said to be
free if it does not occur within the scope of a quantifier. By writing
$\varphi(\vec{x})$, for an MSO formula $\varphi$, and a set of logical
variables $\vec{x}$, we mean that all free variables of $\varphi$ are
in $\vec{x}$.

The syntax of MSO is defined below:
\[\begin{array}{rcl} 
u & \in & PVar;~ x,X ~\in~ LVar_{mso};~k ~\in~ \nat \\
\varphi & ::= & x=y \mid var_u(x) \mid edge_k(x,y) 
\mid null(x) \mid X(x) \mid \varphi \wedge \varphi
\mid \neg\varphi \mid \exists x. \varphi \mid \exists X. \varphi 
\end{array}\]
The semantics of MSO on states is given by the relation $S,\iota,\nu
\models_{mso} \varphi$, where $S = \langle s, h \rangle$ is a state,
$\iota : \{x,y,z,\ldots\} \rightharpoonup_{fin} Loc$ is an
interpretation of the first order variables, and $\nu :
\{X,Y,Z,\ldots\} \rightharpoonup_{fin} 2^{Loc}$ is an interpretation
of the second order variables. If $S,\iota,\nu \models_{mso} \varphi$
for all interpretations $\iota : \{x,y,z,\ldots\}
\rightharpoonup_{fin} Loc$ and $\nu : \{X,Y,Z,\ldots\}
\rightharpoonup_{fin} 2^{Loc}$, then we say that $S$ is a {\em model}
of $\varphi$, denoted $S \models_{mso} \varphi$. We use the standard
MSO semantics \cite{seese91}, with the following interpretations of
the vertex and edge labels:
\[\begin{array}{rcl}
S,\iota,\nu \models_{mso} null(x) & \iff & \iota(x) = nil \\
S,\iota,\nu \models_{mso} var_u(x) & \iff & s(u) =  \iota(x) \\
S,\iota,\nu \models_{mso} edge_k(x,y) & \iff & h_k(\iota(x)) =  \iota(y)
\end{array}\]
The {\em satisfiability problem} for MSO asks, given a formula
$\varphi$, whether there exists a state $S$ such that $S \models_{mso}
\varphi$. This problem is, in general, undecidable. However, one can
show its decidability on a restricted class of models. The theorem
below is a slight variation of a classical result in (MSO-definable)
graph theory \cite{seese91}. For space reasons, all proofs are given
in \cite{TechRep}.

\begin{theorem}\label{mso-bounded-tw}
  Let $k \geq 0$ be an integer constant, and $\varphi$ be an MSO
  formula. The problem asking if there exists a state $S$ such that
  $tw(S) \leq k$ and $S \models_{mso} \varphi$ is decidable.
\end{theorem}

\subsection{Syntax and Semantics of Separation Logic}\label{sec:SepLogic}

Separation Logic (SL) \cite{reynolds02} uses only a set of first order
logical variables, denoted as $LVar_{sl}$, ranging over locations. We
suppose that $LVar_{sl} \cap PVar = \emptyset$ and $LVar_{sl} \cap
LVar_{mso} = \emptyset$. Let $Var_{sl}$ denote the set $PVar \cup
LVar_{sl}$. A formula is said to be {\em closed} if it does not
contain logical variables which are not under the scope of a
quantifier. By writing $\varphi(\vec{x})$ for an SL formula $\varphi$
and a set of logical variables $\vec{x}$, we mean that all free
variables of $\varphi$ are in $\vec{x}$.

\subsubsection{Basic Formulae.}
The syntax of basic formula is given below:
\[\begin{array}{lcl}
  \alpha & \in & Var_{sl} \setminus \{nil\};~ \beta ~\in~ Var_{sl};~ x
  ~\in~ LVar_{sl} \\  \pi & ::= & \alpha = \beta ~|~ \alpha \neq
  \beta ~|~ \pi_1 \wedge \pi_2 \\ \sigma & ::= & emp ~|~ \alpha
  \mapsto (\beta_1,\ldots,\beta_n) ~|~ \sigma_1 * \sigma_2 ~\mbox{,
    for some}~ n > 0 \\ \varphi & ::= & \pi \wedge \sigma ~|~ \exists
  x ~.~ \varphi
\end{array}\]
A formula of the form \mbox{$\bigwedge_{i=1}^n \alpha_i = \beta_i
  ~\wedge~ \bigwedge_{j=1}^m \alpha_j \neq \beta_j$} defined by $\pi$
in the syntax above is said to be \emph{pure}. If $\Pi$ is a pure
formula, let $\Pi^*$ denote its \emph{closure}, i.e., the equivalent
pure formula obtained by the exhaustive application of the
reflexivity, symmetry, and transitivity axioms of equality. A formula
of the form $\bigstar_{i=1}^k \alpha_i \mapsto
(\beta_{i,1},\ldots,\beta_{i,n})$ defined by $\sigma$ in the syntax
above is said to be \emph{spatial}. The atomic proposition $emp$
denotes the empty spatial conjunction. For a spatial formula $\Sigma$,
let $\len{\Sigma}$ be the total number of variable occurrences in
$\Sigma$, e.g. $\len{emp} = 0$, $\len{\alpha \mapsto
  (\beta_1,\ldots,\beta_n)} = n+1$, etc.

The semantics of a basic formula $\varphi$ is given by the relation
$S,\iota \models_{sl} \varphi$ where $S=\langle s, h \rangle$ is a
state, and $\iota : LVar_{sl} \rightharpoonup_{fin} Loc$ is an
interpretation of logical variables from $\varphi$. For a closed
formula $\varphi$, we denote by $S \models_{sl} \varphi$ the fact that
$S$ is a {\em model} of $\varphi$.
\[\begin{array}{lcl}
  S,\iota \models_{sl} emp & \iff & dom(h) = \emptyset \\
  S,\iota \models_{sl} \alpha \mapsto (\beta_1,\ldots,\beta_n) & \iff &
  h = \{\langle (s\oplus\iota)(\alpha), \lambda i ~.~ 
  \mbox{if}~ i \leq n ~\mbox{then}~ (s\oplus\iota)(\beta_i) ~\mbox{else}~ \bot \rangle\} \\
  S,\iota \models_{sl} \varphi_1 * \varphi_2 & \iff & 
  S_1,\iota \models_{sl} \varphi_1 ~\mbox{and}~ S_2,\iota \models_{sl} \varphi_2 
  ~\mbox{where $S_1 \uplus S_2 = S$} \\
\end{array}\]
The semantics of $=$, $\neq$, $\wedge$, and $\exists$ is classical.
Here, the notation $S_1 \uplus S_2 = S$ means that $S$ is the union of
two states $S_1 = \langle s_1,h_1 \rangle$ and $S_2 = \langle s_2,h_2
\rangle$ whose stacks agree on the evaluation of common program
variables ($\forall \alpha \in PVar ~.~ s_1(\alpha) \neq \bot \wedge
s_2(\alpha) \neq \bot \Rightarrow s_1(\alpha) = s_2(\alpha)$), and
whose heaps have disjoint domains ($dom(h_1) \cap dom(h_2) =
\emptyset$) i.e., $S = \langle s_1 \cup s_2, h_1 \oplus h_2
\rangle$. Note that we adopt here the \emph{strict semantics}, in
which a points-to relation $\alpha \mapsto (\beta_1,\ldots,\beta_n)$
holds in a state consisting of a single cell pointed to by $\alpha$,
with exactly $n$ outgoing edges towards dangling locations pointed to
by $\beta_1,\ldots,\beta_n$, and the empty heap is specified by $emp$.

Every basic formula $\varphi$ is equivalent to an existentially
quantified pair $\Sigma \wedge \Pi$ where $\Sigma$ is a spatial
formula and $\Pi$ is a pure formula. Given a basic formula $\varphi$,
one can define its spatial ($\Sigma$) and pure ($\Pi$) parts uniquely,
up to equivalence. A variable $\alpha \in Var$ is said to be
\emph{allocated in $\varphi$} if and only if $\alpha \mapsto (\ldots)$
occurs in $\Sigma$. It is easy to check that an allocated variable may
not refer to a dangling location in any model of $\varphi$. A variable
$\beta$ is \emph{referenced} if and only if $\alpha \mapsto
(\ldots,\beta,\ldots)$ occurs in $\Sigma$ for some variable
$\alpha$. For a basic formula $\varphi \equiv \Sigma \wedge \Pi$, the
{\em size of $\varphi$} is defined as $\len{\varphi} = \len{\Sigma}$.

\begin{lemma}\label{tw-basic}
  Let $\varphi(\vec{x})$ be a basic SL formula, $S = \langle s,h
  \rangle$ be a state, and $\iota : LVar_{sl} \rightharpoonup_{fin}
  Loc$ be an interpretation, such that $S,\iota \models_{sl}
  \varphi(\vec{x})$. Then $tw(S) \leq
  \max(\len{\varphi},\card{PVar})$.
\end{lemma}

\subsubsection{Recursive Definitions.}
A system $\mathcal{P}$ of \emph{recursive definitions} is of the form:
\[\begin{array}{rcl}
P_1(x_{1,1},\ldots,x_{1,n_1}) & ::= & \mid_{j=1}^{m_1}
R_{1,j}(x_{1,1},\ldots,x_{1,n_1}) 
\\ & \ldots & \\ 
P_k(x_{k,1},\ldots,x_{k,n_k}) & ::= & \mid_{j=1}^{m_k} R_{k,j}(x_{k,1},
\ldots, x_{k,n_k}) 
\end{array}\] 
where $P_1,\ldots,P_k$ are called {\em predicates}, $x_{i,1}, \ldots,
x_{i,n_i}$ are called {\em parameters}, and the formulae $R_{i,j}$ are
called the {\em rules} of $P_i$. Concretely, a rule $R_{i,j}$ is of
the form $R_{i,j}(\vec{x}) \equiv \exists \vec{z} ~.~ \Sigma *
P_{i_1}(\vec{y}_1) * \ldots * P_{i_m}(\vec{y}_m) ~\wedge~ \Pi$, where
$\Sigma$ is a spatial SL formula over variables $\vec{x} \cup
\vec{z}$, called the {\em head} of $R_{i,j}$, $\langle
P_{i_1}(\vec{y}_1), \ldots, P_{i_m}(\vec{y}_m) \rangle$ is an ordered
sequence of {\em predicate occurrences}, called the {\em tail} of
$R_{i,j}$ (we assume w.l.o.g. that $\vec{x} \cap \vec{z} = \emptyset$,
and that $\vec{y}_k \subseteq \vec{x} \cup \vec{z}$, for all $k=1,
\ldots, m$), $\Pi$ is a pure formula over variables $\vec{x} \cup
\vec{z}$.

Without losing generality, we assume that all variables occurring in a
rule of a recursive definition system are logical variables from
$LVar_{sl}$ -- pointer variables can be passed as parameters at the
top level. We subsequently denote $head(R_{i,j}) \equiv \Sigma$,
$tail(R_{i,j}) \equiv \langle P_{i_k}(\vec{y}_k) \rangle_{k=1}^m$ and
$pure(R_{i,j}) \equiv \Pi$, for each rule $R_{i,j}$. Rules with empty
tail are called {\em base cases}. For each rule $R_{i,j}$ let
$\card{R_{i,j}}^{var} = \card{\vec{z}} + \card{\vec{x}}$ be the number
of variables, both existentially quantified and parameters, that occur
in $R_{i,j}$. We denote by $\card{\mathcal{P}}^{var} =
\max\{\card{R_{i.j}}^{var} ~|~ 1 \leq i \leq k,~ 1 \leq j \leq m_i\}$
the maximum such number, among all rules in $\mathcal{P}$. We also
denote by $\dir(\mathcal{P}) = \{-1,0, \ldots,
\max\{\len{tail(R_{i,j})} ~|~ 1 \leq i \leq k,~ 1 \leq j \leq m_i\} -
1\}$ the \emph{direction alphabet} of $\mathcal{P}$.

\begin{figure}[h]
\begin{tabular}{ccc}
\begin{minipage}{8.9cm}
\scalebox{0.61}{\begin{tikzpicture}[
level distance=2.5cm,
level 1/.style={sibling distance=7.5cm},
level 2/.style={sibling distance=3.5cm},
bend angle=45,
auto,
every node/.style={transform shape}
]

\node (c1) {
  \shortstack{
    $\exists l^\varepsilon,r^\varepsilon,z^\varepsilon . x^\varepsilon \mapsto
    (l^\varepsilon,r^\varepsilon,p^\varepsilon,nil) \wedge$\\
    $\exists x^0,p^0,leaf_l^0,leaf_r^0,  x^1,p^1,leaf_l^1,leaf_r^1 .~$\\ 
    $  l^\varepsilon = x^0
    \wedge x^\varepsilon = p^0 \wedge leaf_l^\varepsilon = leaf_l^0
    \wedge z^\varepsilon = leaf_r^0 \wedge$\\
   $r^\varepsilon = x^1 \wedge
    x^\varepsilon = p^1 \wedge z^\varepsilon = leaf_l^1 \wedge
    leaf_r^\varepsilon = leaf_r^1$ 
    }

  }
  child {
    node (c11) {
      \shortstack {
        $\exists l^0,r^0,z^0 . x^0 \mapsto (l^0,r^0,p^0,nil)\wedge$\\
        $\exists x^{00},p^{00},leaf_l^{00},leaf_r^{00},  x^{01},p^{01},leaf_l^{01},leaf_r^{01} .~$\\ 
	$l^0 = x^{00} \wedge x^0  = p^{00} \wedge leaf_l^0 = leaf_l^{00} 
	\wedge z^0 = leaf_r^{00} \wedge$ \\
	$r^0 = x^{01} \wedge x^0 = p^{01} \wedge z^0 = leaf_l^{01}
        \wedge leaf_r^0 = leaf_r^{01}$
      }
    }
    child [level distance=2.3cm] {
      node (c111) {
        \shortstack{
		  $x^{00} \mapsto (nil, nil, p^{00}, leaf_r^{00})$\\
		  ${}\wedge x^{00} = leaf_l^{00} $
		}
	  }
    }
    child [level distance=3.2cm] {
      node (c112) {
		\shortstack{
		  $x^{01} \mapsto (nil, nil, p^{01}, leaf_r^{01})$\\
		  ${}\wedge x^{01} = leaf_l^{01} $
		}
	  }
	}
  }
  child {
    node (c12) {
      \shortstack{
        $\exists l^1,r^1,z^1 . x^1 \mapsto (l^1,r^1,p^1,nil)\wedge$\\
        $\exists x^{10},p^{10},leaf_l^{10},leaf_r^{10},  x^{11},p^{11},leaf_l^{11},leaf_r^{11} .~$\\ 
	    $l^1 = x^{10} \wedge x^1 =
	  p^{10} \wedge leaf_l^1 = leaf_l^{10} \wedge z^1 = leaf_r^{10} \wedge$\\
	  $r^1 = x^{11} \wedge x^1 = p^{11} \wedge z^1 = leaf_l^{11}
	  \wedge leaf_r^1 = leaf_r^{11}$
      }
    }
    child [level distance=2.3cm] {
	  node (c121) {
		\shortstack{
		  $x^{10} \mapsto (nil, nil, p^{10}, leaf_r^{10})$\\
		  ${}\wedge x^{10} = leaf_l^{10} $
		}
	  }
	}
    child [level distance=3.2cm] {
	  node (c122) {
		\shortstack{
		  $x^{11} \mapsto (nil, nil, p^{11}, leaf_r^{11})$\\
		  ${}\wedge x^{11} = leaf_l^{11} $
		}
	  }
	}
};

\path
  (c1) -- (c11) node [midway,anchor=center,fill=white] {$*$}
  (c11) -- (c111) node [midway,anchor=center,fill=white] {$*$}
  (c11) -- (c112) node [midway,anchor=center,fill=white] {$*$}
  (c1) -- (c12) node [midway,anchor=center,fill=white] {$*$}
  (c12) -- (c121) node [midway,anchor=center,fill=white] {$*$}
  (c12) -- (c122) node [midway,anchor=center,fill=white] {$*$}
;


\draw[black,<-,dashed] (c121.north) .. controls ($(c1.center)$) .. (c112.north) node [midway,xshift=-1mm,yshift=-4mm] {$z^\varepsilon$};

\draw[black,<-,dashed] 
let \p2=(c112.north), \p1=(c111.north) in
(\x2-.5cm,\y2) .. controls ($(c11.south)!.4!(c11.center)$) .. (\x1+.8cm,\y1) node [midway,xshift=1mm,yshift=-2mm]{$z^0$};

\draw[black,<-,dashed] 
let \p2=(c122.north), \p1=(c121.north) in
(\x2-.5cm,\y2) .. controls ($(c12.south)!.4!(c12.center)$) .. (\x1+.8cm,\y1) node [midway,xshift=1mm,yshift=-2mm]{$z^1$};

\end{tikzpicture}}
\centerline{(a)}
\end{minipage}
&
\begin{minipage}{3cm}
\epsfig{file=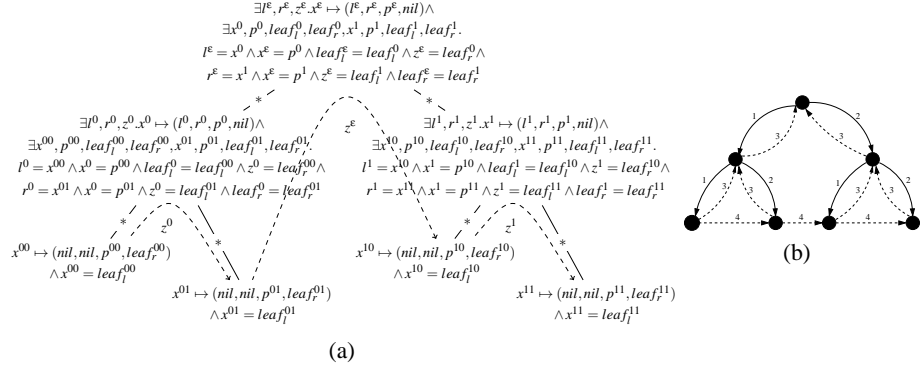,height=18mm}
\centerline{(b)}
\end{minipage}
\end{tabular}
\caption{(a) An unfolding tree for tll predicate and (b) a model of the corresponding formula}
\label{fig:tll-unfolding-tree}
\end{figure}

\paragraph{Example.} 
The predicate $tll$ describes a data structure called a {\em tree with
  parent pointers and linked leaves} (see
Fig. \ref{fig:tll-unfolding-tree}(b)). The data structure is composed of a binary
tree in which each internal node points to left and right children,
and also to its parent node. In addition, the leaves of the tree are
kept in a singly-linked list, according to the order in which they
appear on the frontier (left to right).
\[\begin{array}{rclr}
tll(x,p,leaf_l,leaf_r) & ::= & x \mapsto (nil,nil,p,leaf_r) \wedge x = leaf_l & (R_1) \\ 
& | & \exists l,r,z.~ x \mapsto (l,r,p,nil) * tll(l,x,leaf_l,z) * tll(r,x,z,leaf_r) & (R_2)
\end{array}\]
The base case rule $(R_1)$ allocates leaf nodes. The internal nodes of
the tree are allocated by the rule $(R_2)$, where the $ttl$ predicate
occurs twice, first for the left subtree, and second for the right
subtree. \qed

\begin{definition}\label{unfolding} Given a system of recursive definitions
$\mathcal{P} = \big\{P_i ~::=~ \mid_{j=1}^{m_i}
  R_{i,j}\big\}_{i=1}^{n}$, an \emph{unfolding tree of $\mathcal{P}$
    rooted at $i$} is a finite tree $t$ such that:
  \begin{enumerate}
  \item each node of $t$ is labeled by a single rule of the system $\mathcal{P}$, 
  \item the root of $t$ is labeled with a rule of $P_i$,
  \item nodes labeled with base case rules have no successors, and
  \item if a node $u$ of $t$ is labeled with a rule whose tail is
    $P_{i_1}(\vec{y}_1) * \ldots * P_{i_m}(\vec{y}_m)$, then the
    children of $u$ form the ordered sequence $v_1, \ldots, v_m$ where
    $v_j$ is labeled with one of the rules of $P_{i_j}$ for all
    $j=1,\ldots,m$.
\end{enumerate}
\end{definition}

\paragraph{Remarks.} 
Notice that the recursive predicate $P(x) ::= \exists y ~.~ x \mapsto
y * P(y)$ does not have finite unfolding trees. However, in general a
system of recursive predicates may have infinitely many finite
unfolding trees. \qed

In the following, we denote by $\mathcal{T}_i(\mathcal{P})$ the set of
unfolding trees of $\mathcal{P}$ rooted at $i$. An unfolding tree
$t\in\mathcal{T}_i(\mathcal{P})$ corresponds to a basic formula of
separation logic $\phi_t$, called the {\em characteristic formula} of
$t$, and defined in what follows. For a set of tree positions $P
\subseteq \nat^*$, we denote $LVar^P = \{x^p ~|~ x \in LVar,~ p \in
P\}$. For a tree position $p \in \nat^*$ and a rule $R$, we denote by
$R^p$ the rule obtained by replacing every variable occurrence $x$ in
$R$ by $x^p$. For each position $p \in dom(t)$, we define a formula
$\phi_t^p$, by induction on the structure of the subtree of $t$ rooted
at $p$:
\begin{itemize}
\item if $p$ is a leaf labeled with a base case rule $R$, then
  $\phi_t^p \equiv R^p$
\item if $p$ has successors $p.1,\ldots,p.m$, and the label of $p$ is
  the recursive rule $R(\vec{x}) \equiv \exists \vec{z} ~.~ head(R) *
  \bigstar_{j=1}^m P_{i_j}(\vec{y}_j) \wedge pure(R)$, then:
  $$\phi^p_t(\vec{x}^p) \equiv \exists \vec{z}^p ~.~ head(R^p) *
  \bigstar_{j=1}^m [\exists \vec{x}^{p.i}_{i_j} ~.~
    \phi_t^{p.i}(\vec{x}^{p.i}_{i_j}) \wedge \vec{y}_j^p =
    \vec{x}^{p.i}_{i_j}] \wedge pure(R^p)$$
\end{itemize}
In the rest of the paper, we write $\phi_t$ for
$\phi^\epsilon_t$. Notice that $\phi_t$ is defined using the set of
logical variables $LVar^{dom(t)}$, instead of $LVar$. However the
definition of SL semantics from the previous carries over naturally to
this case.

\paragraph{Example.} (cont'd)  
Fig. \ref{fig:tll-unfolding-tree}(a) presents an unfolding tree for the $tll$
predicate given in the previous example. The characteristic formula of
each node in the tree can be obtained by composing the formulae
labeling the children of the node with the formula labeling the
node. The characteristic formula of the tree is the formula of its
root. \qed

Given a system of recursive definitions $\mathcal{P} = \big\{P_i ~::=~
\mid_{j=1}^{m_i} R_{i,j}\big\}_{i=1}^{n}$, the semantics of a
recursive predicate $P_i$ is defined as follows:
\begin{equation}\label{recursive-semantics}
S,\iota \models_{sl} P_i(x_{i,1},\ldots,x_{i,n_i}) ~\iff~
S,\iota^\epsilon \models_{sl}
\phi_t(x^\epsilon_{i,1},\ldots,x^\epsilon_{i,n_i}), ~\mbox{for some}~
t \in \mathcal{T}_i(\mathcal{P})
\end{equation}
where $\iota^\epsilon(x^\epsilon_{i,j}) \stackrel{def}{=}
\iota(x_{i,j})$ for all $j=1,\ldots,n_i$.

\paragraph{Remark.} Since the recursive predicate $P(x) ::= \exists y ~.~ x \mapsto
y * P(y)$ does not have finite unfolding trees, the formula $\exists x
. P(x)$ is unsatisfiable. \qed

\subsubsection{Top Level Formulae.}
We are now ready to introduce the fragment of {\em Separation Logic
  with Recursive Definitions} (SLRD). A formula in this fragment is an
existentially quantified formula of the following form: $\exists
\vec{z} ~.~ \varphi * P_{i_1} * \ldots * P_{i_n}$, where $\varphi$ is
a basic formula, and $P_{i_j}$ are occurrences of recursive
predicates, with free variables in $PVar \cup \vec{z}$. The semantics
of an SLRD formula is defined in the obvious way, from the semantics
of the basic fragment, and that of the recursive predicates.



\paragraph{Example.} 
The following SLRD formulae, with $PVar=\{root,head\}$, describe both the
set of binary trees with parent pointer and linked leaves, rooted
at $root$, with the leaves linked into a list pointed to by $head$. The difference is that
$\varphi_1$ describes also a tree containing only a single allocated location:
\[\begin{array}{lcl}
\varphi_1 & \equiv & tll(root,nil,head,nil) \\
\varphi_2 & \equiv & \exists l,r,x. root 
\mapsto(l,r,nil,nil)* tll(l,root,head,x)* tll(r,root,x,nil) 
~ \qed \end{array}\] 
We are interested in solving two problems on SLRD formulae, namely
{\em satisfiability} and {\em entailment}. The satisfiability problem
asks, given a closed SLRD formula $\varphi$, whether there exists a
state $S$ such that $S \models_{sl} \varphi$. The entailment problem
asks, given two closed SLRD formulae $\varphi_1$ and $\varphi_2$,
whether for all states $S$, $S \models_{sl} \varphi_1$ implies $S
\models_{sl} \varphi_2$. This is denoted also as $\varphi_1
\models_{sl} \varphi_2$. For instance, in the previous example we have
$\varphi_2 \models_{sl} \varphi_1$, but not $\varphi_1 \models_{sl}
\varphi_2$.

In general, it is possible to reduce an entailment problem $\varphi_1
\models \varphi_2$ to satisfiability of the formula $\varphi_1 \wedge
\neg \varphi_2$. In our case, however, this is not possible directly,
because SLRD is not closed under negation. The decision procedures for
satisfiability and entailment is the subject of the rest of this
paper.

\section{Decidability of Satisfiability and Entailment in SLRD}

The decision procedure for the satisfiability and entailment in SLRD
is based on two ingredients. First, we show that, under certain
natural restrictions on the system of recursive predicates, which
define a fragment of SLRD, called $\mbox{SLRD}_{btw}$, all states that
are models of $\mbox{SLRD}_{btw}$ formulae have {\em bounded tree
  width} (Def. \ref{tw}). These restrictions are as follows:
\begin{enumerate}
\item {\em Progress}: each rule allocates exactly one variable
\item {\em Connectivity}: there is at least one selector edge between
  the variable allocated by a rule and the variable allocated by each
  of its children in the unfolding tree
\item {\em Establishment}: all existentially quantified variables
  in a recursive rule are eventually allocated
\end{enumerate}
Second, we provide a {\em translation of $\mbox{SLRD}_{btw}$ formulae
  into equivalent MSO formulae}, and rely on the fact that
satisfiability of MSO is decidable on classes of states with bounded
tree width.


\subsection{A Decidable Subset of SLRD}

At this point we define the $\mbox{SLRD}_{btw}$ fragment formally, by
defining the three restrictions above. The {\em progress} condition
(1) asks that, for each rule $R$ in the system of recursive
definitions, we have $head(R) \equiv \alpha \mapsto
(\beta_1,\ldots,\beta_n)$, for some variables $\alpha, \beta_1,
\ldots, \beta_n \in Var_{sl}$. The intuition between this restriction
is reflected by the following example.

\paragraph{Example.}
Consider the following system of recursive definitions:
\[ls(x,y) ::=  x \mapsto y \mid 
\exists z,t ~.~ x \mapsto (z,nil) * t \mapsto (nil,y) * ls(z,t)\] The
predicate $ls(x,y)$ defines the set of structures $\{x
(\arrow{1}{})^n z \mapsto t (\arrow{2}{})^n y \mid
n \geq 0\}$, which clearly cannot be defined in MSO. \qed

The {\em connectivity} condition (2) is defined below:
\begin{definition}\label{connected}
A rule $R$ of a system of recursive definitions, such that $head(R)
\equiv \alpha \mapsto(\beta_1, \ldots, \beta_n)$ and $tail(R) \equiv
\langle P_{i_1}(\vec{y}_1), \ldots, P_{i_m}(\vec{y}_m) \rangle$, $m
\geq 1$, is said to be \emph{connected} if and only if the following hold:
\begin{itemize}
\item for each $j=1,\ldots,m$, $(\vec{y}_j)_s = \beta'$, for some $1 \leq s \leq
n_{i_j}$, where $n_{i_j}$ is the number of parameters of $P_{i_j}$
\item $\beta_t = \beta'$ occurs in $pure(R)^*$, for some $1 \leq t
\leq n$ 
\item the $s$-th parameter $x_{i_j,s}$ of $P_{i_j}$ is allocated in
  the heads of all rules of $P_{i_j}$.
\end{itemize}
\end{definition}
In this case we say that between rule $R$ and any rule $Q$ of
$P_{i_j}$, there is a \emph{local edge}, labeled by {\em selector}
$t$. $\mathcal{F}(R,j,Q) \subseteq Sel$ denotes the set of all such
selectors. If all rules of $\mathcal{P}$ are connected, we say that
$\mathcal{P}$ is connected.

\paragraph{Example.} 
The following recursive rule, from the previous $tll$ predicate, is
connected:
$$\exists l,r,z ~.~ x \mapsto (l,r,p,nil) * tll(l,x,leaf_l,z) *
tll(r,x,z,leaf_r) ~(R_2)$$ $R_2$ is connected because the variable $l$
is referenced in $R_2$ and it is passed as the first parameter to
$tll$ in the first recursive call to $tll$. Moreover, the first
parameter ($x$) is allocated by all rules of $tll$. $R_2$ is
connected, for similar reasons. We have $\mathcal{F}(R_2,1,R_2) =
\{1\}$ and $\mathcal{F}(R_2,2,R_2) = \{2\}$. \qed

The {\em establishment} condition (3) is formally defined below. 
\begin{definition}\label{established}
Let $P(x_1,\ldots,x_n) = |_{j=1}^m R_j(x_1,\ldots,x_n)$ be a
predicate in a recursive system of definitions. We say that a
parameter $x_i$, for some $i=1,\ldots,n$ is {\em allocated} in $P$ if
and only if, for all $j=1,\ldots,m$:
\begin{itemize}
\item either $x_i$ is allocated in $head(R_j)$, or 
\item (i) $tail(R_j) = \langle P_{i_1}(\vec{y}_1), \ldots,
  P_{i_k}(\vec{y}_k) \rangle$, (ii) $(\vec{y}_\ell)_s = x_i$ occurs in $pure(R_j)^*$, for some
  $\ell = 1,\ldots,k$, and
  (iii) the $s$-th parameter of $P_{i_\ell}$ is allocated in
  $P_{i_\ell}$ 
\end{itemize}
A system of recursive definitions is
said to be {\em established} if and only if every existentially
quantified variable is allocated.
\end{definition}

\paragraph{Example.}
Let $llextra(x) ::= x \mapsto (nil,nil) ~|~ \exists n,e.~ x \mapsto
(n,e) * llextra(n)$ be a recursive definition system, and let $\phi
::= llextra(head)$, where $head \in PVar$. The models of the formula
$\phi$ are singly-linked lists, where in all locations of the heap,
the first selector points to the next location in the list, and the
second selector is dangling i.e., it can point to any location in the
heap. These dangling selectors may form a squared grid of arbitrary
size, which is a model of the formula $\phi$.  However, the set of
squared grids does not have bounded tree width \cite{seese91}.  The
problem arises due to the existentially quantified variables $e$ which
are never allocated. \qed

Given a system $\mathcal{P}$ of recursive definitions, one can
effectively check whether it is established, by guessing, for each
predicate $P_i(x_{i,1},\ldots,x_{i,n_i})$ of $\mathcal{P}$, the
minimal set of parameters which are allocated in $P_i$, and verify
this guess inductively\footnote{For efficiency, a least fixpoint
  iteration can be used instead of a non-deterministic guess.}. Then,
once the minimal set of allocated parameters is determined for each
predicate, one can check whether every existentially quantified
variable is eventually allocated.

\begin{lemma}\label{tw-pred}
  Let $\mathcal{P} = \{P_i ::= \mid_{j=1}^{m_i}
  R_{ij}(x_{i,1},\ldots,x_{i,n_i})\}_{i=1}^k$ be a established
  system of recursive definitions, and $S = \langle s, h \rangle$ be a
  state, such that $S,\iota \models_{sl}
  P_i(x_{i,1},\ldots,x_{i,n_i})$ for some interpretation $\iota :
  LVar_{sl} \rightharpoonup_{fin} Loc$ and some $1 \leq i \leq
  k$. Then $tw(S) \leq \card{\mathcal{P}}^{var}$.
\end{lemma}
The result of the previous lemma extends to an arbitrary top-level
formula:
\begin{theorem}\label{tw-top}
Let $\mathcal{P} = \{P_i ::= \mid_{j=1}^{m_i}
R_{ij}(x_{i,1},\ldots,x_{i,n_i})\}_{i=1}^k$ be a established
system of recursive definitions, and $S = \langle s, h \rangle$ be a
state, such that $S \models_{sl} \exists \vec{z} ~.~
\varphi(\vec{y}_0) * P_{i_1}(\vec{y}_1) * \ldots *
P_{i_n}(\vec{y}_n)$, where $\varphi$ is a basic SL formula, and
$P_{i_j}$ are predicates of $\mathcal{P}$, and $\vec{y}_i \subseteq
\vec{z}$, for all $i=0,1,\ldots,n$. Then $tw(S) \leq
\max(\card{\vec{z}},\len{\varphi},\card{PVar},\card{\mathcal{P}}^{var})$.
\end{theorem}


\section{From $\mbox{SLRD}_{btw}$ to MSO}

This section describes the translation of a SL formula using
recursively defined predicates into an MSO formula. We denote by
$\Pi(X_0,\ldots,X_i,X)$ the fact that $X_0,\ldots,X_i$ is a partition
of $X$, and by $\Sigma(x,X)$ the fact that $X$ is a singleton with $x$
as the only element. 


\subsection{Converting Basic SL Formulae to MSO}

For every SL logical variable $x \in LVar_{sl}$ we assume the
existence of an MSO logical variable $\overline{x} \in LVar_{mso}$,
which is used to replace $x$ in the translation. For every program
variable $u \in PVar \setminus \{nil\}$ we assume the existence of a
logical variable $\overline{x_u} \in LVar_{mso}$. The special variable
$nil \in LVar_{sl}$ is translated into $\overline{x_{nil}} \in
LVar_{mso}$ (with the associated MSO constraint
$null(\overline{x_{nil}})$). In general, for any pointer or logical
variable $\alpha \in Var_{sl}$, we denote by $\overline{\alpha}$, the
logical MSO variable corresponding to it.

The translation of a pure SL formula $\alpha = \beta$, $\alpha \neq
\beta$, $\pi_1 \wedge \pi_2$ is $\overline{\alpha} =
\overline{\beta}$, $\neg (\overline{\alpha} = \overline{\beta})$,
$\overline{\pi_1} \wedge \overline{\pi_2}$, respectively, where
$\overline{\pi}(\overline{\alpha_1},\ldots,\overline{\alpha_k})$ is
the translation of $\pi(\alpha_1,\ldots,\alpha_k)$. Spatial SL
formulae $\sigma(\alpha_1,\ldots,\alpha_k)$ are translated into MSO
formulae
$\overline{\sigma}(\overline{\alpha_1},\ldots,\overline{\alpha_k},X)$,
where $X$ is used for the set of locations allocated in $\sigma$. The
fact that $X$ actually denotes the domain of the heap, is ensured by
the following MSO constraint:
$$Heap(X) \equiv \forall x \bigvee_{i=1}^{\card{Sel}} (\exists y ~.~
edge_i(x,y)) \leftrightarrow X(x)$$ The translation of basic spatial
formulae is defined by induction on their structure:
\[\begin{array}{lcl}
\overline{emp}(X) & \equiv & \forall x ~.~ \neg X(x) \\ 
\overline{(\alpha \mapsto (\beta_1,\ldots,\beta_n))}(X) & \equiv &
\Sigma(\overline{\alpha},X) ~\wedge~ \bigwedge_{i=1}^n
edge_i(\overline{\alpha},\overline{\beta_i}) ~\wedge~
\bigwedge_{i=n+1}^\card{Sel}\forall x ~.~ \neg edge_i(\overline{\alpha},x) \\ 
\overline{(\sigma_1 * \sigma_2)}(X) & \equiv & \exists Y \exists Z ~.~ \overline{\sigma_1}(Y) 
~\wedge~ \overline{\sigma_2}(Z) ~\wedge~ \Pi(Y,Z,X)
\end{array}\]
The translation of a closed basic SL formula $\varphi$ in
MSO is defined as $\exists X ~.~ \overline{\varphi}(X)$, where
$\overline{\varphi}(X)$ is defined as $\overline{(\pi \wedge
  \sigma)}(X) \equiv \overline{\pi} \wedge \overline{\sigma}(X)$, and
$\overline{(\exists x ~.~ \varphi_1)}(X) \equiv \exists \overline{x}
~.~ \overline{\varphi_1}(X)$. The following lemma proves that the MSO
translation of a basic SL formula defines the same set of models as
the original SL formula.

\begin{lemma}\label{basic-sl-mso}
For any state $S = \langle s, h \rangle$, any interpretation $\iota :
LVar_{sl} \rightharpoonup_{fin} Loc$, and any basic SL formula
$\varphi$, we have $S, \iota \models_{sl} \varphi$ if and only if
$S,\overline{\iota},\nu[X \leftarrow dom(h)] \models_{mso}
\overline{\varphi}(X) ~\wedge~$ $Heap(X)$, where $\overline{\iota} :
LVar_{mso} \rightharpoonup_{fin} Loc$ is an interpretation of first
order variables, such that $\overline{\iota}(x_u) = s(u)$, for all $u
\in PVar$, and $\overline{\iota}(\overline{x}) = \iota(x)$, for all $x
\in LVar_{sl}$, and $\nu : LVar_{mso} \rightharpoonup_{fin} 2^{Loc}$
is any interpretation of second-order variables.
\end{lemma}

\subsection{States and Backbones}
\label{sec:backbone}

The rest of this section is concerned with the MSO definition of
states that are models of recursive SL formulae, i.e. formulae
involving recursively defined predicates. The main idea behind this
encoding is that any part of a state which is the model of a recursive
predicate can be decomposed into a tree-like structure, called the
{\em backbone}, and a set of edges between the nodes in this
tree. Intuitively, the backbone is a spanning tree that uses only
{\em local edges}. For instance, in the state depicted in
Fig. \ref{fig:tll-unfolding-tree}(b), the local edges are drawn in
solid lines.

Let $P_k(x_1, \ldots, x_n)$ be a
recursively defined predicate of a system $\mathcal{P}$, and $S, \iota
\models_{sl} P_k(x_1, \ldots, x_n)$, for some state $S = \langle s, h
\rangle$ and some interpretation $\iota : LVar_{sl} \rightarrow
Loc$. Then $S, \iota \models_{sl} \phi_t$, where $t \in
\mathcal{T}_k(\mathcal{P})$ is an unfolding tree, $\phi_t$ is its
characteristic formula, and $\mu : dom(t) \rightarrow dom(h)$ is the
bijective tree that describes the allocation of nodes in the heap by
rules labeling the unfolding tree.
Recall that the direction alphabet of the system $\mathcal{P}$ is
$\mathcal{D}(\mathcal{P}) = \{-1,0,\ldots,N-1\}$, where $N$ is the
maximum number of predicate occurrences within some rule of
$\mathcal{P}$, and denote $\mathcal{D}_+(\mathcal{P}) =
\mathcal{D}(\mathcal{P}) \setminus \{-1\}$. For each rule $R_{ij}$ in
$\mathcal{P}$ and each direction $d \in \mathcal{D}(\mathcal{P})$, we
introduce a second order variable $X_{ij}^d$ to denote the set of
locations $\ell$ such that (i) $t(\mu^{-1}(\ell))\equiv R_{ij}$ and
(ii) $\mu^{-1}(\ell)$ is a $d$-th child, if $d\geq0$, or
$\mu^{-1}(\ell)$ is the root of $t$, if $d=-1$.  Let $\seq{X}$ be the
sequence of $X_{ij}^k$ variables, enumerated in some order. We use the
following shorthands:
\[\begin{array}{lcc}
X_{ij}(x) & \equiv &
\displaystyle\bigvee_{k\in\mathcal{D}(\mathcal{P})}X_{ij}^k(x) \\
\end{array}\hspace*{1cm}
\begin{array}{lcc}
X_{i}(x) & \equiv & \displaystyle\bigvee_{1\leq j\leq m_i}X_{ij}(x) \\
\end{array}\hspace*{1cm}
\begin{array}{lcc}
X_{i}^k(x) & \equiv & \displaystyle\bigvee_{1\leq j\leq m_i}X_{ij}^k(x)
\end{array}\]
to denote, respectively, locations that are allocated by a rule
$R_{ij}$ ($X_{ij}$), by a recursive predicate $P_i$ ($X_i$), or by a
predicate $P_i$, who are mapped to a $k$-th child (or to the root, if
$k=-1$) in the unfolding tree of $\mathcal{P}$, rooted at $i$
($X_i^k$).

In order to characterize the backbone of a state, one must first
define the local edges:
\[\begin{array}{ccl}
local\_edge^d_{i,j,p,q}(x,y) & \equiv &  
\bigwedge_{s \in \mathcal{F}(R_{i,j},d,R_{pq})} edge_s(x,y) 
\end{array}\]
for all $d \in \mathcal{D}_+(\mathcal{P})$. Here
$\mathcal{F}(R_{ij},d,R_{pq})$ is the set of forward local selectors
for direction $d$, which was defined previously -- notice that the set
of local edges depends on the source and destination rules $R_{ij}$
and $R_{pq}$, that label the corresponding nodes in the unfolding
tree, respectively. The following predicate ensures that these labels
are used correctly, and define the successor functions in the
unfolding tree:
\[\begin{array}{rccl}
succ_d(x,y,\seq{X}) & \equiv & \bigvee & 
X_{ij}(x) ~\wedge~ X_{pq}^k(y) ~\wedge~ local\_edge^d_{i,j,p,q}(x,y) \\
&& {\scriptsize\begin{array}{rcl}
     1 & \leq i,p \leq & M \\
     1 & \leq j \leq & m_i \\
     1 & \leq q \leq & m_p
   \end{array}}
\end{array}\]
for all $d \in \mathcal{D}_+(\mathcal{P})$. The definition of the
backbone of a recursive predicate $P_i$ in MSO follows tightly the
definition of the unfolding tree of $\mathcal{P}$ rooted at $i$
(Def. \ref{unfolding}):
$$backbone_i(r,\seq{X},T) \equiv tree(r,\seq{X},T) ~\wedge~
X_i^{-1}(r) ~\wedge~ succ\_labels(\seq{X})$$ where $tree(r,\seq{X},T)$
defines a tree\footnote{
\ifLongVersion
For space reasons this definition is deferred
to Appendix \ref{app:tree-mso}.
\else  
For space reasons this definition can be fund in \cite{TechRep}.
\fi
} 
with domain $T$, rooted at $r$,
with successor functions defined by $succ_0,\ldots,succ_{N-1}$, and
$succ\_labels$ ensures that the labeling of each tree position (with
rules of $\mathcal{P}$) is consistent with the definition of
$\mathcal{P}$:
\[\begin{array}{lcccl}
succ\_labels(\seq{X}) & \equiv & \bigwedge & X_{ij}(x) \rightarrow & 
\bigwedge_{d=0}^{r_{ij}-1} \exists y ~.~ X_{k_d}^d(y) \wedge succ_d(x,y,\seq{X}) 
\\ 
&& {\scriptsize\begin{array}{lcr} 1 & \leq i \leq & M \\ 1 & \leq j \leq & m_i \end{array}} &&
\wedge~ \forall y ~.~ \bigwedge_{p = s_{ij}+1}^{\card{Sel}} \neg edge_p(x,y)
\end{array}\]
where we suppose that, for each rule $R_{ij}$ of $\mathcal{P}$, we
have $head(R_{ij}) \equiv \alpha \mapsto
(\beta_1,\ldots,\beta_{s_{ij}})$ and $tail(R_{ij}) = \langle P_{k_1},
\ldots, P_{k_{r_{ij}}} \rangle$, for some $r_{ij} \geq 0$, and some
indexing $k_1,\ldots,k_{r_{ij}}$ of predicate occurrences within
$R_{ij}$. The last conjunct ensures that a location allocated in
$R_{ij}$ does not have more outgoing edges than specified by
$head(R_{ij})$. This condition is needed, since, unlike SL, the
semantics of MSO does not impose strictness conditions on the number
of outgoing edges.


\subsection{Inner Edges}
\label{sec:non-local-edges}

An edge between two locations is said to be {\em inner} if both
locations are allocated in the heap. 
Let $\mu$ be the bijective tree defined in Sec. \ref{sec:backbone}.
The existence of an edge $\ell \arrow{k}{} \ell'$ in $S$, between two
arbitrary locations $\ell,\ell' \in dom(h)$, is the consequence of:
\begin{enumerate}
\item a basic points-to formula $\alpha \mapsto (\beta_1, \ldots,
  \beta_k, \ldots, \beta_n)$ that occurs in $\mu(\ell)$
\item a basic points-to formula $\gamma \mapsto ( \ldots )$ that
  occurs in $\mu(\ell')$
\item a path $\mu(\ell) = p_1, p_2, \ldots, p_{m-1}, p_m = \mu(\ell')$
  in $t$, such that the equalities $\beta_k^{p_1} = \delta_2^{p_2} =
  \ldots = \delta_{m-1}^{p_{m-1}} = \gamma^{p_m}$ are all logical
  consequences of $\phi_t$, for \ some \ tree positions
  $p_2, \ldots, p_{m-1} \in dom(t)$ and some variables
  $\delta_2,\ldots,\delta_{m-1} \in LVar_{sl}$.
\end{enumerate}
Notice that the above conditions hold only for inner edges. The
(corner) case of edges leading to dangling locations is dealt with in
\ifLongVersion
Appendix \ref{app:parameters}.
\else
\cite{TechRep}.
\fi

\paragraph{Example.} 
The existence of the edge from tree position $00$ to $01$ in
Fig. \ref{fig:tll-unfolding-tree}(b), is a consequence of the
following: (1) $x^{00}\mapsto(nil,nil,p^{00},leaf_r^{00})$, (2)
$x^{01}\mapsto(nil,nil,p^{01},leaf_r^{01})$, and (3)
$leaf_r^{00}=z^0=leaf_l^{01}=x^{01}$.
The reason for other dashed edges is similar.\qed

The main idea here is to encode in MSO the existence of such paths, in
the unfolding tree, between the source and the destination of an edge,
and use this encoding to define the edges. To this end, we use a
special class of tree automata, called {\em tree-walking automata}
(TWA) to recognize paths corresponding to sequences of equalities
occurring within characteristic formulae of unfolding trees.

\paragraph{\bf Tree Walking Automata}
Given a set of tree directions $\mathcal{D} = \{-1,0,\ldots,N\}$ for
some $N\geq0$, a tree-walking automaton\footnote{This notion of
  tree-walking automaton is a slightly modified but equivalent to the
  one in \cite{bojanczyk}. We give the translation of TWA into the
  original definition in 
\ifLongVersion
  Appendix \ref{app:stwa}.
\else
  \cite{TechRep}
\fi
  }, is a tuple $A =
(\Sigma, Q, q_i, q_f, \Delta)$ where $\Sigma$ is a set of tree node
labels, $Q$ is a set of states, $q_i, q_f \in Q$ are the initial and
final states, and $\Delta : Q \times (\Sigma \cup \{root\}) \times
(\Sigma \cup \{?\}) \rightarrow 2^{Q ~\times~ (\mathcal{D} ~\cup~
  \{\epsilon\})}$ is the (non-deterministic) transition function. A
configuration of $A$ is a pair $\langle p, q \rangle$, where $p \in
\mathcal{D}^*$ is a tree position, and $q \in Q$ is a state. A run of
$A$ over a $\Sigma$-labeled tree $t$ is a sequence of configurations
$\langle p_1, q_1 \rangle, \ldots, \langle p_n, q_n \rangle$, with
$p_1, \ldots, p_n \in dom(t)$, such that for all $i=1, \ldots, n-1$,
we have $p_{i+1} = p_i.k$, where either:
\begin{enumerate}
\item $p_i \neq \epsilon$ and $(q_{i+1},k) \in
  \Delta(q_i,t(p_i),t(p_i.(-1)))$, for $k \in \mathcal{D} \cup
  \{\epsilon\}$
  \item $p_i = \epsilon$ and $(q_{i+1},k) \in \Delta(q_i,\sigma,?)$,
    for $\sigma \in \{t(p_i) \cup root\}$ and $k \in \mathcal{D}
    \cup \{\epsilon\}$
\end{enumerate}
The run is said to be {\em accepting} if $q_1 = q_i$, $p_1 = \epsilon$
and $q_n = q_f$. 

\paragraph{\bf Routing Automata}
For a system of recursive definitions $\mathcal{P} =
\big\{P_i(x_{i,1},\ldots,x_{i,n_i}) ::= |_{j=1}^{m_i} R_{ij}
(x_{i,1},\ldots,x_{i,n_i})\big\}_{i=1}^k$, we define the TWA
$A_{\mathcal{P}} = (\Sigma_{\mathcal{P}}, Q_{\mathcal{P}}, q_i, q_f,
\Delta_{\mathcal{P}})$, where $\Sigma_{\mathcal{P}} = \{R_{ij}^k ~|~ 1
\leq i \leq k,~ 1 \leq j \leq m_i, ~ k\in\mathcal{D}(\mathcal{P})\}$, 
$Q_{\mathcal{P}} = \{q^{var}_x ~|~ x \in LVar_{sl}\} \cup \{q^{sel}_s
~|~ s \in Sel\} \cup \{q_{i},q_{f}\}$. The transition function
$\Delta_{\mathcal{P}}$ is defined as follows:
\begin{enumerate}
\item $(q_{i}, k), (q^{sel}_s,\epsilon) \in \Delta(q_{i},\sigma,\tau)$
  for all $k \in \mathcal{D}_+(\mathcal{P})$, all $s \in Sel$ and all
  $\sigma \in \Sigma_{\mathcal{P}} \cup \{root\}$,
  $\tau \in \Sigma_{\mathcal{P}} \cup \{?\}$ i.e., the automaton first
  moves downwards chosing random directions, while in $q_{i}$, then
  changes to $q^{sel}_s$ for some non-deterministically chosen
  selector $s$.

\item $(q^{var}_{\beta_s}, \epsilon) \in \Delta(q^{sel}_s, R_{ij}^k,
  \tau)$ and $(q_{f},\epsilon) \in \Delta(q^{var}_\alpha,
  R_{ij}^k, \tau)$ for all $k\in\mathcal{D}(\mathcal{P})$ and
  $\tau \in \Sigma_{\mathcal{P}} \cup \{?\}$ if and only if
  $head(R_{ij}) \equiv \alpha \mapsto
  (\beta_1, \ldots, \beta_s, \ldots, \beta_m)$, for some $m > 0$ i.e.,
  when in $q^{sel}_s$, the automaton starts tracking the destination
  $\beta_s$ of the selector $s$ through the tree. The automaton enters
  the final state when the tracked variable $\alpha$ is allocated.

\item for all $k \in \mathcal{D}_+(\mathcal{P})$, all $\ell \in \mathcal{D}(\mathcal{P})$ 
      and all rules $R_{\ell q}$ of $P_\ell(x_{\ell,1}, \ldots,
  x_{\ell, n_\ell})$, we have $(q^{var}_{x_{\ell,j}},
  k) \in \Delta(q^{var}_{y_j}, R_{ij}^l, \tau)$, for all
  $\tau \in \Sigma_{\mathcal{P}} \cup \{?\}$, and $(q^{var}_{y_j},
  -1) \in \Delta(q^{var}_{x_{\ell,j}}, R_{\ell q}^k, R_{ij}^l)$ if and
  only if $tail(R_{ij})_k \equiv P_\ell(y_1, \ldots, y_{n_\ell})$
  i.e., the automaton moves down along the $k$-th direction tracking
  $x_{\ell,j}$ instead of $y_j$, when the predicate $P_\ell(\vec{y})$
  occurs on the $k$-th position in $R_{ij}$. Symmetrically, the
  automaton can also move up tracking $y_j$ instead of $x_{\ell,j}$,
  in the same conditions.

\item  
  $(q^{var}_\beta,\epsilon) \in \Delta(q^{var}_\alpha, R_{ij}^k,\tau)$
  for all $k \in \mathcal{D}(\mathcal{P})$ and all
  $\tau \in \Sigma_{\mathcal{P}} \cup \{?\}$ if and only if
  $\alpha=\beta$ occurs in $pure(R_{ij})$ i.e., the automaton switches
  from tracking $\alpha$ to tracking $\beta$ when the equality between
  the two variables occurs in $R_{ij}$, while keeping the same
  position in the tree.
\end{enumerate}
The following lemma formalizes the correctness of the TWA construction:
\begin{lemma}\label{twa-run}
  Given a system of recursive definitions $\mathcal{P}$, and an
  unfolding tree $t \in \mathcal{T}_i(\mathcal{P})$ of $\mathcal{P}$,
  rooted at $i$, for any $x,y \in LVar_{sl}$ and $p,r \in dom(t)$, we
  have $\models_{sl} \phi_t \rightarrow x^p = y^r$ if and only if
  $A_{\mathcal{P}}$ has a run from $\langle p,q^{var}_x \rangle$ to
  $\langle r,q^{var}_y \rangle$ over $t$, where $\phi_t$ is the
  characteristic formula of $t$.
\end{lemma}

To the routing automaton $A_{\mathcal{P}}$ corresponds the MSO formula
$\Phi_{A_{\mathcal{P}}}(r,\seq{X},T,\seq{Y})$, where $r$ maps to the
root of the unfolding tree, $\seq{X}$ is the sequence of second order
variables $X_{ij}^k$ defined previously, $T$ maps to the domain of the
tree, and $\seq{Y}$ is a sequence of second-order variables $X_q$, one
for each state $q \in Q_{\mathcal{P}}$. We denote by $Y^{sel}_s$ and
$Y_f$ the variables from $\seq{Y}$ that correspond to the states
$q^{sel}_S$ and $q_f$, for all $s \in Sel$, respectively. For space
reasons, the definition of $\Phi_{A_{\mathcal{P}}}$ is given in
\ifLongVersion
Appendix \ref{app:twa-mso}.
\else
\cite{TechRep}.
\fi
With this notation, we define:
$$inner\_edges(r,\seq{X},T) \equiv \forall x \forall y \bigwedge_{s
  \in Sel} \exists \seq{Y} ~.~
\Phi_{A_{\mathcal{P}}}(r,\seq{X},T,\seq{Y}) \wedge Y^{sel}_s(x) \wedge
Y_f(y) \rightarrow edge_s(x,y)$$

\subsection{Double Allocation}\label{app:double-alloc}

In order to translate the definition of a recursively defined SL
predicate $P(x_1,\ldots,x_n)$ into an MSO formula $\overline{P}$, that
captures the models of $P$, we need to introduce a sanity condition,
imposing that recursive predicates which establish equalities between
variables allocated at different positions in the unfolding tree, are
unsatisfiable, due to the semantics of the separating conjunction of
SL, which implicitly conjoins all local formulae of an unfolding tree.
%
%
A double allocation occurs in the unfolding tree $t$ if and only if
there exist two distinct positions $p,q \in dom(t)$ and:
\begin{enumerate}
\item a basic points-to formula $\alpha \mapsto (\ldots)$ occurring in $t(p)$
\item a basic points-to formula $\beta \mapsto (\ldots)$ occurring in $t(q)$
\item a path $p=p_1,\ldots,p_m=q$ in $t$, such that the equalities
  $\alpha^p = \gamma_2^{p_2} = \ldots = \gamma_{m-1}^{p_{m-1}} =
  \beta^q$ are all logical consequences of $\phi_t$, for some tree
  positions $p_2,\ldots,p_{m-1} \in dom(t)$ and some variables
  $\gamma_2,\ldots,\gamma_{m-1} \in LVar_{sl}$
\end{enumerate}
The cases of double allocation can be recognized using a routing
automaton $B_{\mathcal{P}} = (\Sigma_{\mathcal{P}}, Q'_{\mathcal{P}},
q_i, q_f, \Delta'_{\mathcal{P}})$, whose states $Q'_{\mathcal{P}} =
\{q^{var}_x ~|~ x \in LVar_{sl}\} \cup \{q_0,q_i, q_f\}$ and
transitions $\Delta'_{\mathcal{P}}$ differ from $A_{\mathcal{P}}$ only
in the following rules:
\begin{itemize}
\item $(q_0,\epsilon) \in \Delta(q_i,\sigma,\tau)$ for all
  $\sigma \in \Sigma_{\mathcal{P}} \cup \{root\}$ and all $\tau \in
  \Sigma_{\mathcal{P}} \cup \{?\}$, i.e. after non-deterministically
  chosing a position in the tree, the automaton enters a designated
  state $q_0$, which occurs only once in each run.
\item $(q^{var}_\alpha,\epsilon) \in \Delta(q_0,R_{ij}^k,\tau)$ for
  all $k\in\mathcal{D}(\mathcal{P})$ and all $\tau \in
  \Sigma_{\mathcal{P}} \cup \{?\}$ \ if  and \ only  if $head(R_{ij}) =
  \alpha \mapsto (\ldots)$, while in the designated state $q_0$, the
  automaton starts tracking the variable $\alpha$, which is allocated
  at that position.
\end{itemize}
This routing automaton has a run over $t$, which labels one position
by $q_0$ and a distinct one by $q_f$ if and only if two positions in
$t$ allocate the same location. Notice that $B_{\mathcal{P}}$ has
always a trivial run that starts and ends in the same position --
since each position $p \in dom(t)$ allocates a variable $\alpha$, and
$\langle q_i, \epsilon \rangle, \ldots, \langle q_0, p \rangle,
\langle q^{var}_\alpha, p \rangle, \langle q_f, p \rangle$ is a valid
run of $B_{\mathcal{P}}$. The predicate system has no double
allocation if and only if these are the only possible runs of
$B_{\mathcal{P}}$.

The existence of a run of $B_{\mathcal{P}}$ is captured by an MSO
formula $\Phi_{B_\mathcal{P}}(r,\seq{X},T,\seq{Y})$, where $r$ maps to
the root of the unfolding tree, $\seq{X}$ is the sequence of second
order variables $X_{ij}^k$ defined previously, $T$ maps to the domain
of the tree, and $\seq{Y}$ is the sequence of second-order variables
$Y_q$, taken in some order, each of which maps to the set of tree
positions visited by the automaton while in state $q \in
Q'_{\mathcal{P}}$ -- we denote by $Y_0$ and $Y_f$ the variables from
$\seq{Y}$ that correspond to the states $q_0$ and $q_f$,
respectively. Finally, we define the constraint:
$no\_double\_alloc(r,\seq{X},T) \equiv \forall \seq{Y} ~.~ 
\Phi_{B_{\mathcal{P}}}(r,\seq{X},T,\seq{Y}) \rightarrow Y_0=Y_f$

\subsection{Handling Parameters}\label{app:parameters}

The last issue to be dealt with is the role of the actual parameters
passed to a recursively defined predicate $P_i(x_{i,1}, \ldots,
x_{i,k})$ of $\mathcal{P}$, in a top-level formula. Then, for each
parameter $x_{i,j}$ of $P_i$ and each unfolding tree
$t \in \mathcal{T}_i(\mathcal{P})$, there exists a path $\epsilon =
p_1,\ldots,p_m \in dom(t)$ and variables $\alpha_1,\ldots,\alpha_m \in
LVar_{sl}$ such that $x_{i,j} \equiv
\alpha_1$ and $\alpha_\ell^{p_\ell} = \alpha_{\ell+1}^{p_{\ell+1}}$ is a
consequence of $\phi_t$, for all $\ell=1,\ldots,m-1$. Subsequently,
 there are three (not necessarily disjoint)  possibilities: 
\begin{enumerate} 
\item $head(t(p_m)) \equiv \alpha_m \mapsto (\ldots)$, i.e. $\alpha_m$ 
      is allocated
\item $head(t(p_m)) \equiv \beta \mapsto(\gamma_1, \ldots, \gamma_p,
      \ldots, \gamma_\ell)$, and $\alpha_m \equiv \gamma_p$,
      i.e. $\alpha_m$ is referenced
\item $\alpha_m \equiv x_{i,q}$ and $p_m = \epsilon$, for some 
$1 \leq q \leq k$, i.e. $\alpha_m$ is another parameter $x_{i,q}$
\end{enumerate} 
Again, we use slightly modified routing automata (one for each of the
 case above) $C^{i,j}_{\mathcal{P},c} = (\Sigma_{\mathcal{P}},
 Q''_{\mathcal{P}}, q_i, q_f, \Delta_c^{i,j})$ for the cases $c =
 1,2,3$, respectively. Here $Q''_{\mathcal{P}} = \{q^{var}_x ~|~
 x \in LVar_{sl} \} \cup \{q_s^{sel} \mid s \in Sel \} \cup 
\{q^{i,a} \mid 1 \leq a \leq k \} \cup \{q_i,q_f\}$ and $\Delta^{i,j}_c$, 
$c=1, 2, 3$ differ from the transitions of $A_{\mathcal{P}}$ in the
following:
\begin{itemize}
\item $(q^{i,j},\epsilon) \in \Delta_x^{i,j}(q_i,root,?)$, i.e. the
  automaton marks the root of the tree with a designated state
  $q^{i,j}$, that occurs only once on each run

\item $(q^{var}_{x_{i,j}},\epsilon) \in \Delta_x^{i,j}(q^{i,j},R_{ik}^{-1},?)$,
  for each rule $R_{ik}$ of $P_i$, i.e. the automaton starts tracking
  the parameter variable $x_{i,j}$ beginning with the root of the tree

\item $(q_f,\epsilon) \in \Delta_1^{i,j}(q^{var}_\alpha, R_{ij}^k, \tau)$, 
  for all $k\in\mathcal{D}(\mathcal{P})$, $\tau \in \Sigma_{\mathcal{P}} \cup \{?\}$ 
  iff $head(R_{ij}) \equiv \alpha \mapsto (\dots)$ is the final rule for $C^{i,j}_{\mathcal{P},1}$

\item $(q_s^{sel},\epsilon) \in \Delta_2^{i,j}(q^{var}_\gamma, R_{ij}^k, \tau)$, 
  for all $k\in\mathcal{D}(\mathcal{P})$ and
  $\tau \in \Sigma_{\mathcal{P}} \cup \{?\}$ iff
  $head(R_{ij}) \equiv \alpha \mapsto
  (\beta_1, \ldots, \beta_s, \ldots, \beta_n)$ and
  $\gamma \equiv \beta_s$ i.e., $q_s^{sel}$ is reached in the second
  case, when the tracked variable is referenced. After that, 
  $C^{i,j}_{\mathcal{P},2}$ moves to the final state i.e.,
  $(q_f,\epsilon)\in \Delta_2^{i,j}(q_s^{sel}, \sigma, \tau)$ for all
  $s\in Sel$, all $\sigma \in \Sigma_{\mathcal{P}} \cup \{root\}$ and
  $\tau \in \Sigma_{\mathcal{P}} \cup \{?\}$

\item $(q^{i,a},\epsilon) \in \Delta_3^{i,j}(q^{var}_{x_{i,a}},root,?)$ and 
  $(q_f,\epsilon)\in \Delta_3^{i,j}(q_{i,a},root , ?)$, for each
  $1 \leq a\leq k$ and $a\neq j$ i.e., are the final moves for 
  $C^{i,j}_{\mathcal{P},3}$
\end{itemize}
The outcome of this construction are MSO formulae
$\Phi_{C^{i,j}_{\mathcal{P},c}}(r,\seq{X},T,\seq{Y})$, for $c=1,2,3$,
where $r$ maps to the root of the unfolding tree, respectively,
$\seq{X}$ is the sequence of second order variables $X_{ij}^k$ defined
previously, $T$ maps to the domain of the tree, and $\seq{Y}$ is the
sequence of second order variables corresponding to states of
$Q''_{\mathcal{P}}$ -- we denote by $Y_f,
Y^{i,a},Y_s^{sel} \in \seq{Y}$ the variables corresponding to the
states $q_f$, $q^{i,a}$, and $q_s^{sel}$, respectively. The parameter
$x_{i,j}$ of $P_i$ is assigned by the following MSO constraints:
\[\begin{array}{rcl}
param^1_{i,j}(r,\seq{X},T) & \equiv & \exists \seq{Y} ~.~
\Phi_{C^{i,j}_{\mathcal{P},1}} ~\wedge~
Y^{i,j}_0(\overline{x}_{i,j}) ~\wedge~ \forall y ~.~ Y_f(y)
\rightarrow \overline{x}_{i,j} = y
\\
param^2_{i,j}(r,\seq{X},T) & \equiv & \exists \seq{Y} ~.~
\Phi_{C^{i,j}_{\mathcal{P},2}} ~\wedge~
Y^{i,j}_0(\overline{x}_{i,j}) ~\wedge~ 
\bigwedge_{s\in Sel} 
\forall y ~.~ Y_s^{sel}(y) \rightarrow edge_s(y,\overline{x}_{i,j})
\\
param^3_{i,j}(r,\seq{X},T) & \equiv & \exists \seq{Y} ~.~
\Phi_{C^{i,j}_{\mathcal{P},3}} ~\wedge~
Y^{i,j}_0(\overline{x}_{i,j}) ~\wedge~ 
\bigwedge_{1 \leq a\leq k}
\forall y ~.~ Y^{i,a}(y) \rightarrow \overline{x}_{i,j}= \overline{x}_{i,a}
\end{array}\]
where $\overline{x}_{i,j}$ is the first-order MSO variable
corresponding to the SL parameter $x_{i,j}$. Finally, the constraint
$param_{i,j}$ is conjunction of the $param_{i,j}^c,~ c=1,2,3$ formulae.

\subsection{Translating Top Level $\mbox{SLRD}_{btw}$ Formulae to MSO}

We define the MSO formula corresponding to a predicate $P_i(x_{i,1},
\ldots, x_{i,n_i})$, of a system of recursive definitions $\mathcal{P}
= \{P_1, \ldots, P_n\}$:
\[\begin{array}{rcl}
\overline{P_i}(\overline{x}_{i,1}, \ldots, \overline{x}_{i,n_i},T) &
\equiv & \exists r \exists \seq{X} ~.~ backbone_i(r,\seq{X},T)
~\wedge~ inner\_edges(r,\seq{X},T) ~\wedge 
\\ 
&& no\_double\_alloc(r,\seq{X},T) ~\wedge~ \bigwedge_{1 \leq j \leq n_i}
param_{i,j}(r,\seq{X},T)
\end{array}\]
The following lemma is needed to establish the correctness of our
construction.
\begin{lemma}\label{predicate-sl-mso} 
  For any state $S=\langle s,h \rangle$, any interpretation $\iota :
  LVar_{sl} \rightarrow_{fin} Loc$, and any recursively defined
  predicate $P_i(x_1,\ldots,x_n)$, we have $S,\iota \models_{sl}
  P_i(x_1,\dots,x_n)$ if and only if $S,\overline{\iota},\nu[T
    \leftarrow dom(h)] \models_{mso}
  \overline{P_i}(\overline{x_1},\ldots,\overline{x_k},T) \wedge
  Heap(T)$, where $\overline{\iota} : LVar_{mso} \rightharpoonup_{fin}
  Loc$ is an interpretation of first order variables, such that
  $\overline{\iota}(x_u) = s(u)$, for all $u \in PVar$, and
  $\overline{\iota}(\overline{x}) = \iota(x)$, for all $x \in
  LVar_{sl}$, and $\nu : LVar_{mso} \rightharpoonup_{fin} 2^{Loc}$ is
  any interpretation of second-order variables.
\end{lemma}
Recall that a top level $\mbox{SLRD}_{btw}$ formula is of the form: $\varphi \equiv
\exists \vec{z} ~.~ \phi(\vec{y}_0) * P_{i_1}(\vec{y}_1) * \ldots
P_{i_k}(\vec{y}_k)$, where $1 \leq i_1, \ldots, i_k \leq n$, and
$\vec{y}_j \subseteq \vec{z}$, for all $j=0,1,\ldots,k$. We define the
MSO formula: $$\overline{\varphi}(X) \equiv \exists \overline{\vec{z}}
\exists X_{0,\ldots,k} ~.~ \overline{\phi}(\overline{\vec{y}_0},X_0)
~\wedge~ \overline{P_{i_1}}(\overline{\vec{y}_1},X_1) ~\wedge~ \ldots
~\wedge~ \overline{P_{i_k}}(\overline{\vec{y}_k},X_k) ~\wedge~
\Pi(X_0,X_1,\ldots,X_k,X)$$

\begin{theorem}\label{slrd-mso}
  For any state $S$ and any closed $\mbox{SLRD}_{btw}$ formula $\varphi$ we have that
  $S \models_{sl} \varphi$ if and only if $S \models_{mso} \exists X
  ~.~ \overline{\varphi}(X) ~\wedge~ Heap(X)$.
\end{theorem}
Theorem \ref{tw-top} and the above theorem prove decidability of
satisfiability and entailment problems for $\mbox{SLRD}_{btw}$, by
reduction to MSO over states of bounded tree width.

\section{Conclusions and Future Work}

We defined a fragment of Separation Logic with Recursive Definitions,
capable of describing general unbounded mutable data structures, such
as trees with parent pointers and linked leaves. The logic is shown to
be decidable for satisfiability and entailment, by reduction to MSO
over graphs of bounded tree width. We conjecture that the complexity
of the decision problems for this logic is elementary, and plan to
compute tight upper bounds, in the near future.

\paragraph*{Acknowledgement.} This work was supported by the Czech Science
Foundation (project P103/10/0306) and French National Research Agency
(project VERIDYC ANR-09-SEGI-016). We also acknowledge Tom\'a\v{s}
Vojnar, Luk\'a\v{s} Hol\'ik and the anonymous reviewers for their
valuable comments.

\bibliographystyle{splncs03}
\bibliography{refs}

\ifLongVersion
\else
\end{document}
\eject
\fi

\newpage
\appendix

\section{Definition of tree structures in MSO}\label{app:tree-mso}
Let $\seq{X} = \{X_1,\ldots,X_m\}$ define a set of tree labels. Given
a direction alphabet $\mathcal{D} = \{-1,0,\ldots,N\}$, we consider a
set of (partial) successor functions $Succ_{\mathcal{D}} =
\{succ_0,$ $\ldots,succ_N\}$. These functions can be encoded by MSO
formulae $succ_d(x,y,\seq{X})$, and are supposed to satisfy the
following constraint: $$\forall x,y,z ~.~ \bigwedge_{i \in
  \mathcal{D}_+} succ_i(x,y) \wedge succ_i(x,z) \rightarrow y=z$$ A
tree structure with root $x$, domain $X$ and successor functions
$Succ_{\mathcal{D}}$ is defined by the MSO formula
$tree(r,\seq{X},T)$, which is the conjunction of $\Pi(\seq{X},T)$ and
the following four MSO constraints:
\begin{enumerate}[(A)]
\item $x$ is the root of the tree:
  $$\forall y ~.~ X(y) \rightarrow \bigwedge_{d \in \mathcal{D}_+}
  \neg succ_d(y,x,\seq{X})$$
\item successors are pairwise distinct:
  $$\forall x,y,z ~.~ \bigwedge_{0 \leq i < j \leq N}
  succ_i(x,y,\seq{X}) \wedge succ_j(x,z,\seq{X}) \rightarrow y \neq
  z$$
\item each node except for the root has exactly one predecessor:
  $$\forall y ~.~ x \neq y \rightarrow \exists! z ~.~ \bigvee_{i \in
  \mathcal{D}_+} succ_i(z,y)$$ where $\exists!$ stands for the unique
  existential quantification
\item all nodes in $X$ are reachable from $x$:
  \[\begin{array}{rcl}
  closed(X) & \equiv & \forall y ~.~ X(y) \wedge \bigwedge_{i \in
    \mathcal{D}_+} \exists z ~.~ succ_i(y,z) \rightarrow X(z)
  \\ reach(x,X) & \equiv & X(x) \wedge closed(X) \wedge \forall Y ~.~
  Y(x) \wedge closed(Y) \rightarrow X \subseteq Y
  \end{array}\]
\end{enumerate}
The following lemma formalizes the correctness of this definition.
\begin{lemma}\label{tree-mso}
  For any state $S = \langle s, h \rangle$ and interpretations $\iota
  : LVar_{mso} \rightharpoonup_{fin} Loc$ and $\nu : LVar_{mso}
  \rightharpoonup_{fin} 2^{Loc}$ such that $\iota(r) \in dom(h)$ and
  $\nu(X_i) \subseteq dom(h)$, for all $i=1,\ldots,m$, and $\nu(T) =
  dom(h)$, we have: $$S,\iota,\nu \models_{mso} tree(r,\seq{X},T)$$ if
  and only if there exists a unique prefix-closed set $P \subseteq
  \nat^*$ and two unique trees $\mu : P \rightarrow dom(h)$ and
  $\lambda : P \rightarrow \{X_1,\ldots,X_m\}$, such that $\mu$ is
  bijective, $\mu(\epsilon) = \iota(r)$, and:
  \begin{enumerate}
  \item $\forall \ell \in dom(h)~ \forall d \in \mathcal{D}_+ ~.~
    S,\iota[x \leftarrow \ell][y \leftarrow \mu(\epsilon)],\nu 
    \models_{mso} \neg succ_d(x,y,\seq{X})$
  \item $\forall p \in P~ \forall d \in \mathcal{D}_+ ~.~ S,\iota[x
    \leftarrow \mu(p)][y \leftarrow \mu(p.d)],\nu \models_{mso}
    succ_d(x,y,\seq{X}) \iff p.d \in D$
  \item $\nu(X_i) = \{\ell \in dom(h) ~|~ \lambda(\mu^{-1}(\ell)) = X_i\}$
  \end{enumerate}
\end{lemma}
\proof{``$\Rightarrow$'' Since $S,\iota,\nu \models_{mso}
  \Pi(\seq{X},T)$, we have that $\nu(X_1),\ldots,\nu(X_m)$ forms a
  partition of $dom(h)$. We define $D$, $\mu$ and $\lambda$, as the
  limits of the increasing sequences defined as follows. Let $P_0 =
  \{\epsilon\}$, $\mu_0 = \{(\epsilon, \iota(r))\}$ and $\lambda_0 =
  \{(\epsilon,X_j)\}$, where $j\in\{1,\ldots,m\}$ is the unique index
  such that $\iota(r) \in \nu(X_j)$, and, for all $i \geq 0$:
  \begin{itemize}
  \item $P_{i+1} = P_i \cup \{p.d ~|~ p \in P_i, d \in \mathcal{D}_+,
    p.d \not\in P_i,~ \exists \ell \in dom(h) ~.~ S,\iota[x \leftarrow
      \mu_i(p)][y \leftarrow \ell],\nu \models_{mso} succ_d(x,y,\seq{X})\}$
      
  \item $\mu_{i+1} = \mu_i \cup \{(p.d,\ell) ~|~ p \in D_i, d \in
    \mathcal{D}_+, p.d \not\in P_i,~ S,\iota[x \leftarrow \mu_i(p)][y
      \leftarrow \ell],\nu \models_{mso} succ_d(x,y,\seq{X}) \}$;
    notice that the choice of $\ell$ is unique, because $succ_d$
    defines a partial function
    
  \item $\lambda_{i+1} = \lambda_i \cup \{(p,X_j) ~|~ p \in
    dom(\mu_{i+1}) \setminus dom(\mu_i),~ \mu_{i+1}(p) \in
    \nu(X_j)\}$, where $j\in\{1,\ldots,m\}$ is the unique index such
    that $\mu_{i+1}(p) \in \nu(X_j)$
  \end{itemize}
  The sequences stabilize, because $dom(h)$ is finite, and we define
  $P = \bigcup_{i\geq 0} P_i$, $\mu = \bigcup_{i\geq 0} \mu_i$ and
  $\lambda = \bigcup_{i\geq 0} \lambda_i$. Moreover $\mu_0(\epsilon) =
  \iota(r)$, hence $\mu(\epsilon) = \iota(r)$. The first condition
  is satisfied as a consequence of point (A) in the definition of
  $tree(r,\seq{X},T)$, and the second condition can be proved by
  induction on the definitions of $P_i$ and $\mu_i$. The fact that
  $\mu$ is bijective is a consequence of points (B), (C) and (D) in
  the definition of $tree(r,\seq{X},T)$. First, suppose that $\mu$ is
  not one-to-one, i.e. there exist two distinct positions $p,q\in
  dom(\mu)$ such that $\mu(p)=\mu(q)=\ell$. Since $p\neq q$, either:
  \begin{enumerate}[(i)]
    \item $p$ is a prefix of $q$, or viceversa
    \item there exist a position $r$ such that $r.d_1$ is a prefix of
      $p$ and $r.d_2$ is a prefix of $q$, for some $d_1,d_2 \in
      \mathcal{D}_+$, $d_1 \neq d_2$.
  \end{enumerate}
  In both cases we obtain a contradiction. Second, suppose that $\mu$
  is not onto, i.e. there exists $\ell \in dom(h)$ such that $\mu(p)
  \neq \ell$, for all $p \in dom(\mu)$. But this is clearly in
  contradiction with point (D) above and the definition of $D$ and
  $\mu$. The third condition can be proved inductively on the
  definition of $\lambda_i$. Finally $P$, $\mu$ and $\lambda$ are
  unique, since the choices at each step $i \geq 0$ in the definition
  of $P_i$, $\mu_i$ and $\lambda_i$ are unique. ``$\Leftarrow$'' This
  direction is an easy exercise. \qed}









\section{Standard Tree Walking Automata}\label{app:stwa}

We recall the standard definition of a TWA from \cite{bojanczyk}.
Given a set of tree directions $\mathcal{D} = \{-1,0,\ldots,N\}$, for
some $N\geq0$, a {\em standard tree-walking automaton} (STWA) is a
tuple $A_s = (\Sigma, Q, q_i, q_f, \Delta_s)$ where $\Sigma$ is a set
of tree node labels, $Q$ is a set of states, $q_i,q_f \in Q$ are the
initial and final states, and $\Delta_s : Q \times (\mathcal{D}_+
~\cup~ \{root\}) \times \Sigma \rightarrow 2^{Q ~\times~ (\mathcal{D}
  \cup \{\epsilon\})}$ is the (non-deterministic)
transition function. A configuration of $A$ is a pair $\langle p, q
\rangle$, where $p \in \mathcal{D}_+^*$ is a tree position, and $q \in
Q$ is a state. A run of $A$ over a $\Sigma$-labeled tree $t$ is a
sequence of configurations $\langle p_1, q_1 \rangle, \ldots, \langle
p_n, q_n \rangle$, with $p_1, \ldots, p_n \in dom(t)$ and
$q_1,\ldots,q_n \in Q$, such that, for all $i=1, \ldots, n-1$, we have
$p_{i+1} = p_i.k$ for some $k \in \mathcal{D} \cup \{\epsilon\}$,
where either:
\begin{enumerate}
\item $p_i = p.d$ for some $p \in \mathcal{D}_+^*$, $d \in
  \mathcal{D}_+$ and $(q_{i+1},k) \in \Delta(q_i,d,t(p_i))$
\item $p_i = \epsilon$ and $(q_{i+1},k) \in \Delta(q_i,root,t(p_i))$
\end{enumerate}
The run is said to be {\em accepting} if $q_1 = q_i$, $p_1 = \epsilon$
and $q_n = q_f$, in which case we say that $A$ accepts $t$. We denote
by $\lang{A}$ the set of trees accepted by a (S)TWA $A$.

\begin{lemma}\label{twa-stwa}
For each TWA $A=(\Sigma, Q, q_i, q_f, \Delta)$ there exists an STWA \\
$A_s=(\Sigma,Q \cup Q_{aux}, q_i,q_f, \Delta_s)$ such that
$\lang{A} = \lang{A_s}$.
\end{lemma}
\proof{For each rule in $\Delta$ we create a set of rules in
  $\Delta_s$, such that $\langle p_1,q_1 \rangle
  \stackrel{A}{\rightarrow} \langle p_2,q_2 \rangle$ if and only if
  $\langle p_1,q_1 \rangle \stackrel{A_s}{\rightarrow}^+ \langle
  p_2,q_2 \rangle$, i.e. we simulate the effect of a single step in
  $A$ by a sequence of steps in $A_s$. The construction of $\Delta_s$
  is done as follows. Let $(q_j,k) \in \Delta(q_i, \sigma, \pi)$ be a
  transition rule of $A$.
\begin{itemize}
\item if $\sigma=root$ and $\pi=?$ then, for each $\tau \in \Sigma$,
  we have $(q_j,k) \in \Delta_s(q_i, root, \tau)$
\item if $\sigma \in \Sigma$ and $\pi=?$ then $(q_j,k) \in
  \Delta_s(q_i, root, \sigma)$
\item if $\sigma,\pi \in \Sigma$ then, for each $d \in \mathcal{D}_+$,
  we have the following sequence of rules:
  \begin{itemize}
  \item $(q_{i,d}^1,-1) \in \Delta_s(q_i, d, \sigma)$
  \item $(q_{i,d}^2, d) \in \Delta_s(q_{i,d}^1, e, \pi)$, for each $e \in
    \mathcal{D}_+ \cup \{root\}$
  \item $(q_j,k) \in \Delta_s(q_{i,d}^2, d, \sigma)$
  \end{itemize}
  for two fresh states $q_{i,m}^1, q_{i,m}^2 \in Q_{aux} \setminus Q$.


\end{itemize} 
The following proofs are left as an easy exercise. 
\begin{enumerate}
\item each sequence of steps $\langle p_1, q_1 \rangle, \ldots,
  \langle p_n, q_n \rangle$ of $A$ corresponds to a unique sequence of
  steps of $A_s$ starting and ending in the same configurations
\item each sequence of steps $\langle p_1, q_1 \rangle, \ldots,
  \langle p_n, q_n \rangle$ of $A_s$ corresponds to a unique sequence of
  steps of $A$ starting and ending in the same configurations
\end{enumerate}
\qed}

\section{MSO encoding of Tree Walking Automata}\label{app:twa-mso}

We consider a class of tree structures with successor functions
$Succ_{\mathcal{D}} = \{succ_i ~|~ i \in \mathcal{D}_+\}$ for some set
of directions $\mathcal{D} = \{-1,0,\ldots,N\}$, $N \geq 0$, with
labels from the alphabet $\seq{X} = \langle X_1, \ldots, X_m \rangle$
of second-order variables. These labels define a partition on the
domain of the tree, i.e. we assume that the following constraint holds
in what follows: $$tree(r,\seq{X},T)$$ We define the predecessor
function $succ_{-1}$ as follows:
$$succ_{-1}(x,y,\seq{X}) \equiv \bigvee_{0\leq i \leq k}
succ_i(y,x,\seq{X})$$ Since the successors of a node in a tree are
pairwise distinct, $succ_{-1}$ is a well-defined partial function.

Let $A=(\seq{X},Q,q_{i},q_{f},\Delta)$ be a tree walking automaton, and
let $Q = \{q_1,\ldots,q_k\}$ be some arbitrary indexing of the set of
states. W.l.o.g. we assume that no transition rule in $\Delta$
originates in $q_f$. Let $\seq{Y} = \langle Y_1,\ldots,Y_k \rangle$ be
a sequence of second-order variables, one for each state. First, we
define a step $\langle x, q_i \rangle \rightarrow \langle y, q_j
\rangle$ of $A$ on the tree, as follows:
\[\begin{array}{rccl}
step(x, y, \seq{X}) & \equiv & \bigvee & Y_i(x) ~\wedge~ Y_j(y) ~\wedge~ X_p(x)
~\wedge~ succ_k(x,y,\seq{X}) ~\wedge \\ && {\scriptscriptstyle (q_j, k) \in
  \Delta(q_i,X_p,X_q)} & \big(x = r ~~\vee~ \exists z ~.~
succ_{-1}(x,z,\seq{X}) \wedge X_q(z)\big)
\end{array}\]
Any position on the run is reachable from the root $r$, with respect
to the $step$ relation:
\[\begin{array}{rcl}
step\_closed(X) & \equiv & \forall x,y ~.~ X(x) \wedge step(x,y,\seq{X}) 
\rightarrow X(y) 
\\ 
step\_reach(x,X) & \equiv & X(x) \wedge step\_closed(X) \wedge 
\forall Y ~.~ Y(x) \wedge step\_closed(Y) \rightarrow X \subseteq Y
\end{array}\]
The run $R$ of $A$ is defined by the conjunction of the following
constraints:
\begin{enumerate}[(A)]
\item $R$ equals the union of $Y_1,\ldots,Y_k$ i.e., each position in
  the run is marked by at least one state of the automaton
\item the root $r$ of the tree is labeled with $Y_i$, where $Y_i$ is
  the second-order variable corresponding to the initial state $q_i$:
  $$Y_i(r)$$
\item the final position of $x_f$ the run is labeled with $Y_f$, where $Y_f$
  is the second-order variable corresponding to the final state
  $q_i$: $$\forall y ~.~ \neg step(x_f,y,\seq{X}) \wedge Y_f(x_f)$$
\item every non-final position has a successor position in the run: 
  $$R \subseteq T \wedge step\_reach(r,R) \wedge \forall
  x ~.~ x \neq x_f \rightarrow \exists y ~.~ step(x,y,\seq{X})$$
\end{enumerate}
The final formula $\Phi_A(r,\seq{X},T,\seq{Y})$ is obtained by
conjoining the above constraints and existentially quantifying
$x_f$ and $R$. The following lemma formalizes the correctness of this
construction:
\begin{lemma}\label{twa-mso}
For any state $S = \langle s, h \rangle$ and interpretations $\iota :
LVar_{mso} \rightharpoonup_{fin} Loc$ and $\nu : LVar_{mso}
\rightharpoonup_{fin} 2^{Loc}$ where $\iota(r) \in dom(h)$ and
$\nu(X_i) \subseteq dom(h)$, for all $i = 1,\ldots,m$ and $\nu(T) =
dom(h)$, such that: $$S,\iota,\nu \models_{mso} tree(r,\seq{X},T)$$
let $P \subseteq \nat^*$ and $\mu : P \rightarrow dom(h)$, $\lambda :
P \rightarrow \{X_1,\ldots.X_m\}$ be the prefix-closed set and unique
trees from Lemma \ref{tree-mso}. Then we have: $$S,\iota,\nu
\models_{mso} \Phi_{A}(r,\seq{X},T,\seq{Y})$$ if and only if $A$ has a
loop-free accepting run $\pi$ over $\lambda$ such that $\nu(Y_j) =
\{\mu(p) ~|~ p \in P ~\mbox{and $\langle p, q_j \rangle$ occurs on
  $\pi$}\}$, for all $j=1,\ldots,k$.
\end{lemma}
\proof{``$\Rightarrow$'' From the definition of $\Phi_A$, we can
  construct a loop-free maximal path $\ell_0, \ell_1, \ldots, \ell_n$
  in $dom(h)$ such that:
  \begin{itemize}
  \item $\iota(r) = \ell_0 \in \nu(Y_i)$ and $\ell_n \in \nu(Y_f)$
  \item $S,\iota,\nu[x \leftarrow \ell_i][y \leftarrow \ell_{i+1}]
    \models_{mso} step(x,y,\seq{X})$, for all $i \geq 0$
  \item $\bigcup_{j=1}^k \nu(Y_j) = \{\ell_0,\ell_1,\ldots,\ell_n\}$
  \end{itemize}
  Then $\epsilon = \mu^{-1}(\ell_0), \mu^{-1}(\ell_1), \ldots,
  \mu^{-1}(\ell_n)$ is a path in $P$ and $A$ has a loop-free accepting
  run $\pi : \langle \mu^{-1}(\ell_0), q_i \rangle, \ldots, \langle
  \mu^{-1}(\ell_n), q_f \rangle$ over $\lambda$. Moreover $\langle
  \mu^{-1}(\ell_i), q_j \rangle$ occurs on the run if and only if
  $\ell_i \in \nu(Y_j)$. ``$\Leftarrow$'' This direction is left as an
  easy exercise. \qed}

\section{Routing Automaton Example} 

Routing automaton for the tree with link leaves---predicate $tll$ from 
Section \ref{sec:SepLogic}. There is only a
single predicate $tll$ with two rules. The routing automaton is
$A_{tll}=(\Sigma_{\mathcal{P}},Q,q_i,q_f,\Delta)$, where
\begin{itemize}
\item $\Sigma_{\mathcal{P}}  =  \{R_{11}^0,R_{12}^0,R_{11}^1,R_{12}^1,R_{11}^{-1},R_{12}^{-1}  \}$
\item $Q  =   \{q_x^{var},q_p^{var},q_{leaf_l}^{var},q_{leaf_r}^{var},q_l^{var},q_r^{var},q_z^{var} \} \cup
\{q^{sel}_1,q^{sel}_2,q^{sel}_3,q^{sel}_4\} \cup \{q_i,q_f\}$
\end{itemize}

 and $\Delta$ is defined as follows (the numbers corresponds to numbers in definition of
 routing automata):
\begin{enumerate}
\item
\begin{itemize}
\item
$(q_i,k),(q^{sel}_1,\epsilon),(q^{sel}_2,\epsilon),(q^{sel}_3,\epsilon),(q^{sel}_4,\epsilon)
\in \Delta(q_i,\sigma,\tau),\sigma\in\Sigma_{\mathcal{P}}, k\in \{0,1\},
\tau\in\Sigma_{\mathcal{P}}\cup \{?\}$
\end{itemize}

\item
\begin{itemize}
\item $(q_l^{var},\epsilon)\in \Delta_{init}(q^{sel}_1,R_{12}^k,
\tau),k\in\mathcal{D}(\mathcal{P}),\tau\in\Sigma_{\mathcal{P}}\cup \{?\} $
\item $(q_r^{var} ,\epsilon)\in \Delta_{init}(q^{sel}_2,R_{12}^k,
\tau),k\in\mathcal{D}(\mathcal{P}),\tau\in\Sigma_{\mathcal{P}}\cup \{?\} $
\item $(q_p^{var} ,\epsilon)\in \Delta_{init}(q^{sel}_3,\sigma,\tau),
\sigma\in\Sigma_{\mathcal{P}},\tau\in\Sigma_{\mathcal{P}}\cup \{?\} $
\item $(q_{leaf_r}^{var} ,\epsilon)\in \Delta_{init}(q^{sel}_4,R_{11}^k, \tau),k\in\mathcal{D}(\mathcal{P}), \tau\in\Sigma_{\mathcal{P}}\cup \{?\}$
\item $(q_f,\epsilon)\in \Delta( q_x^{var}, \sigma, \tau),\sigma\in\Sigma_{\mathcal{P}},
\tau\in\Sigma_{\mathcal{P}}\cup \{?\}$
\end{itemize}

\item
\begin{itemize}
\item 
$(q_x^{var} ,0) \in \Delta (q_l^{var},R_{12}^k,\tau),k\in\mathcal{D}(\mathcal{P}),\tau\in\Sigma_{\mathcal{P}}\cup \{?\}$
\item 
$(q_l^{var} ,-1)  \in \Delta (q_x^{var} ,\sigma^0,\tau),\sigma\in\{R_{11},R_{12}\},\tau\in\Sigma_{\mathcal{P}}$
\item $ 
(q_x^{var} ,1) \in \Delta ( q_r^{var} ,R_{12}^k,\tau),k\in\mathcal{D}(\mathcal{P}),\tau\in\Sigma_{\mathcal{P}}\cup \{?\}$
\item $ 
(q_r^{var} ,-1)  \in \Delta (q_x^{var}
,\sigma^1,\tau),\sigma\in\{R_{11},R_{12}\},\tau\in\Sigma_{\mathcal{P}}$

\item $ 
(q_p^{var},0),(q_p^{var},1) \in \Delta ( q_x^{var} , R_{12}^k,\tau),k\in\mathcal{D}(\mathcal{P}),\tau\in\Sigma_{\mathcal{P}}\cup \{?\}$
\item $ 
(q_x^{var} ,-1)   \in \Delta (q_p^{var},\sigma,\tau),\sigma\in\Sigma_{\mathcal{P}},\tau\in\Sigma_{\mathcal{P}} $

\item $ 
(q_{leaf_r}^{var} ,0),(q_{leaf_l}^{var} ,1) \in \Delta (q_z^{var} ,R_{12}^k,\tau),k\in\mathcal{D}(\mathcal{P}),\tau\in\Sigma_{\mathcal{P}}\cup \{?\}$

\item $ 
(q_z^{var} ,-1)  \in \Delta (
q_{leaf_l}^{var} ,\sigma^1,\tau),\sigma\in\{R_{11},R_{12}\},\tau\in\Sigma_{\mathcal{P}}$

\item $ 
(q_z^{var} ,-1)  \in \Delta (q_{leaf_r}^{var} ,\sigma^0,\tau),\sigma\in\{R_{11},R_{12}\},\tau\in\Sigma_{\mathcal{P}}$

\item $ 
(q_{leaf_l}^{var} ,0) \in \Delta (q_{leaf_l}^{var} ,R_{12}^k,\tau),k\in\mathcal{D}(\mathcal{P}),\tau\in\Sigma_{\mathcal{P}}\cup \{?\}$

\item $ 
(q_{leaf_l}^{var} , -1)  \in \Delta (
q_{leaf_l}^{var} , \sigma^0,\tau),\sigma\in\{R_{11},R_{12}\},\tau\in\Sigma_{\mathcal{P}}$

\item $ 
(q_{leaf_r}^{var},1) \in \Delta
(q_{leaf_r}^{var},R_{12}^k,\tau),k\in\mathcal{D}(\mathcal{P}),\tau\in\Sigma_{\mathcal{P}}\cup \{?\}$

\item $ 
(q_{leaf_r}^{var} ,-1)  \in \Delta (q_{leaf_r}^{var} ,\sigma^1,\tau),\sigma\in\{R_{11},R_{12}\},\tau\in\Sigma_{\mathcal{P}}$

\end{itemize}

\item
\begin{itemize}
\item $(q_x^{var},\epsilon) \in \Delta (q_{leaf_l}^{var},R_{11}^k,\tau),k\in\mathcal{D}(\mathcal{P}),\tau\in\Sigma_{\mathcal{P}}\cup \{?\}$
\item $(q_{leaf_l}^{var} ,\epsilon) \in \Delta (q_x^{var},R_{11}^k,\tau),k\in\mathcal{D}(\mathcal{P}),\tau\in\Sigma_{\mathcal{P}}\cup \{?\}$
\end{itemize}

\end{enumerate}

\section{Missing Proofs}\label{app:proofs}

{\bf Proof of Theorem \ref{mso-bounded-tw}:} Each state $S = \langle
s, h \rangle$ is both a vertex- and edge-labeled graph, in whose set
of vertices is $loc(S)$, vertices are labeled with pointer variables,
and edges with selectors. There are three MSO-definable restrictions
making the difference betweena states and arbitrary graphs:
\begin{enumerate}
\item each pointer variable $u \in PVar$ labels at most one
  location: $$\rho_1 \equiv \forall x.y ~.~ \bigwedge_{u \in PVar}
  var_u(x) \wedge var_u(y) \rightarrow x=y$$
\item each edge leads to at most one location:
  $$\rho_2 \equiv \forall x,y,z ~.~ \bigwedge_{s \in Sel}
  edge_s(x,y) \wedge edge_s(x,z) \rightarrow y=z$$
\item there exists a unique designated nil location with no
  outgoing edges: $$\rho_3 \equiv \exists! x \forall y ~.~
  \bigwedge_{s \in Sel} \neg edge_s(x,y) \wedge \forall z ~.~
  null(y) \rightarrow x = z$$
\end{enumerate}
Hence the satisfiability problem for an MSO formula $\varphi$
interpreted over states is equivalent to the satisfiability of the MSO
formula $\varphi \wedge \rho_1 \wedge \rho_2 \wedge \rho_3$
interpreted over arbitrary graphs. The latter problem is decidable, as
shown, e.g. by Theorem 2.1 in \cite{madhusudan-parlato11}. \qed

\vspace*{\baselineskip}\noindent {\bf Proof of Lemma \ref{tw-basic}:}
Let $\varphi \equiv \Sigma \wedge \Pi$, where $\Sigma$ and $\Pi$ are
the spatial and pure parts of $\varphi$, respectivelly.  First observe
that $dom(h) = \{(s \oplus \iota)(\alpha) ~|~ \mbox{$\alpha \mapsto
  (\ldots,\beta,\ldots)$ occurs in $\Sigma$}\}$ and $Img(h) = \{(s
\oplus \iota)(\beta) ~|~ \mbox{$\alpha \mapsto (\ldots,\beta,\ldots)$
  occurs in $\Sigma$}\}$. Hence $\len{\varphi} = \len{\Sigma} \geq
\card{dom(h) \cup Img(h)}$. Moreover, we have that $\card{img(s)} \leq
\card{PVar}$.

  We define a tree decomposition of $S$ as follows. Let $P \subseteq
  \nat^*$ be a prefix-closed set such that $\card{P} = \card{dom(h)
    \cup Img(h)}$ and $\delta : \{0.p ~|~ p \in P\} \cup \{\epsilon\}
  \rightarrow 2^{loc(S)}$ be a tree such that $\delta(0.p) = dom(h) \cup
  Img(h)$, for all $p \in P$ and $\delta(\epsilon) = img(s) \setminus
  (dom(h) \cup Img(h))$. It is easy to check that $\delta$ satisfies
  the conditions of Def. \ref{tw}. Also, $\card{\delta(0.p)} \leq
  \len{\varphi}$, for all $p \in P$, and $\card{\delta(\epsilon)} \leq
  \card{PVar}$ i.e., $\card{\delta(p)} \leq
  \max(\len{\varphi},\card{PVar})$, for all $p \in dom(\delta)$. \qed

\vspace*{\baselineskip}\noindent {\bf Proof of Lemma \ref{tw-pred}:}
By the definition of the semantics of recursive predicates, we
have: $$S,\iota^\epsilon \models_{sl} \phi_t(x^\epsilon_{i,1}, \ldots,
x^\epsilon_{i,n_i})$$ for some unfolding tree $t \in
\mathcal{T}_i(\mathcal{P})$. Observe that the only free variables of
$\phi_t$ are $x^\epsilon_{i,1}, \ldots, x^\epsilon_{i,n_i}$, the rest
occurring under existential quantification. Let $\overline{\phi_t}$ be
the (matrix) formula obtained from $\phi_t$ by renaming each
existentially quantified variable to a unique name, and forgetting the
existential quantifiers. Also let $\overline{\iota^\epsilon} :
LVar^{dom(t)} \rightharpoonup_{fin} Loc$ be an interpretation such
that:
  $$S, \overline{\iota^\epsilon} \models_{sl} \overline{\phi_t}$$ By
the definition of the semantics of SL, such an interpretation must
exist. Hence for each position $p \in dom(t)$, there exists a state
$S_p = \langle s, h_p \rangle$ such that:
  $$S_p,\overline{\iota^\epsilon} \models_{sl} head(t(p))$$ and,
moreover $\card{dom(h_p)}=1$, since, by convention, $head(t(p))$
allocates exactly one variable. Consequently, there exists a bijective
tree $\mu : dom(t) \rightarrow dom(h)$ such that, for all $p \in
dom(t)$, we have $dom(h_p) = \{\mu(p)\}$.

We define a tree decomposition $\delta : dom(t) \rightarrow 2^{Loc}$
as follows, for all positions $p_0 \in dom(t)$, $\delta(p_0)$ contains
only the following locations:
  \begin{enumerate}[(i)]
  \item $\mu(p_0) \in \delta(p_0)$
  \item $\iota(x_{i,j}) \in \delta(p_0)$, for all $1 \leq j \leq n_i$
  \item if $head(t(p_0)) \equiv \alpha \mapsto
    (\beta_1,\ldots,\beta_s)$, for each $u = 1,\ldots,s$ and each
    sequence $\beta_u^{p_0} = \ldots = \gamma^{p_1}$ of equalities
    occurring in $\phi_t$, such that $head(t(p_1))$ allocates
    $\gamma$, we have $\mu(p_1) \in \delta(p)$, for each position $p$
    within the sequence
  \end{enumerate}
  First, we prove that $\delta$ is a valid tree decomposition
  (Def. \ref{tw}):
  \begin{enumerate}
  \item Let $\ell \in loc(S)$ be a location. If $\ell \in dom(h)$,
    then $\ell \in \delta(\mu^{-1}(\ell))$ by point (i) above. If
    $\ell \in Img(h) \setminus dom(h)$, since $\mathcal{P}$ is
    established, then $\ell = \iota(x_{i,j})$, for some $1 \leq j
    \leq n_i$, by point (ii) above. Hence $\ell \in \delta(p)$, for
    all $p \in dom(\delta)$. Consequently $loc(S) \subseteq \bigcup_{p
      \in dom(\delta)} \delta(p)$, and the other direction is trivial.

  \item Let $\ell_1 \arrow{s}{} \ell_2$ be an edge in $S$. Then
    $\ell_1 \in dom(h)$ and $\ell_1 \in \delta(\mu^{-1}(\ell_1))$, by
    point (i) above. But $\ell_1 \arrow{s}{} \ell_2$ only exists
    because $head(t(\mu^{-1}(\ell_1))) \equiv \alpha \mapsto
    (\beta_1,\ldots,\beta_s)$, and there exists a sequence
    $\beta_u^{~\mu^{-1}(\ell_1)} = \ldots =
    \gamma^{~\mu^{-1}(\ell_2)}$ of equalities occurring in $\phi_t$,
    for some $u = 1,\ldots,s$, such that $head(t(\mu^{-1}(\ell_2)))$
    allocates $\gamma$. By point (iii) above, we have $\ell_2 \in
    \delta(\mu^{-1}(\ell_1))$.

  \item Let $p,r \in dom(\delta)$ be two distinct locations, $q$ be on
    the path from $p$ to $r$, and let $\ell \in \delta(p) \cap
    \delta(r)$. Then there are two possibilities. Either $\ell =
    \iota(x_{i,j})$, for some $1 \leq j \leq n_i$, in which case $\ell
    \in \delta(q)$, by point (ii) above. Otherwise, the only remaining
    possibility is that both $p$ and $r$ are on a sequence of
    equalities $\alpha_1^{p_1} = \ldots = \alpha_m^{p_m}$, and $\ell
    \in \bigcap_{i=1}^m\delta(p_i)$, cf. point (iii) above. But in
    this case $q$ must be one the same sequence, hence $\ell \in
    \delta(q)$.
  \end{enumerate}
  Finally, we prove that $\card{\delta(p)} \leq
  \card{\mathcal{P}}^{var}$, for all $p \in dom(\delta)$. Let $\ell
  \in \delta(p)$ be a location. There are two reasons for $\ell \in
  \delta(p)$:
  \begin{itemize}
  \item $\ell = \iota(x_{i,j})$, for some $j=1,\ldots,n_i$, cf. point
    (ii) above
  \item $\ell = \overline{\iota^\epsilon}(\alpha^p)$, where $\alpha$
    is an existentially quantified variable that occurs within $t(p)$,
    cf. points (i) and (iii) above
  \end{itemize}
  Hence $\card{\delta(p)}$ may not exceed the maximum number of
  variables that occur either free, or existentially quantified,
  within $t(p)$. \qed

\vspace*{\baselineskip}\noindent {\bf Proof of Theorem \ref{tw-top}:}
Let $\iota : LVar_{sl} \rightharpoonup_{fin} Loc$ be an
interpretation, and $S_0 = \langle s_0, h_0 \rangle, S_1 = \langle
s_1, h_1 \rangle , \ldots, S_n = \langle s_n, h_n \rangle$ be states
such that:
  \[\begin{array}{rcl}
  S_0, \iota & \models_{sl} & \varphi \\
  S_1, \iota & \models_{sl} & P_{i_1} \\
  & \ldots & \\
  S_n, \iota & \models_{sl} & P_{i_n}
  \end{array}\]
  and $S=S_0 \uplus S_1 \uplus \ldots \uplus S_n$. By Lemma
  \ref{tw-basic}, $tw(S_0) \leq \max(\len{\varphi},\card{PVar})$, and
  by Lemma \ref{tw-pred}, $tw(S_i) \leq \card{\mathcal{P}}^{var}$, for
  all $i = 1, \ldots, n$. Hence there exist prefix-closed sets $P_i$
  and tree decompositions $\delta_i : P_i \rightarrow loc(S_i)$ of
  $S_i$, for all $i=0,1,\ldots,n$, respectivelly. 

  We define the prefix-closed set $P = \{i.p ~|~ i=0,1,\ldots,n,~ p
  \in P_i\} \cup \{\epsilon\}$ and a tree decomposition $\delta$ of
  $S$ as follows.  Let $\delta(\epsilon) = \{\iota(x) ~|~ x \in
  \vec{z}\}$ and $\delta(i.p) = \delta_i(p)$, for all $p \in P_i$, and
  $i=0,1,\ldots,n$. Let us first check that $\delta$ meets the
  conditions of Def. \ref{tw}. The first point follows from the fact
  that $loc(S) = \bigcup_{i=0}^n loc(S_i)$, and $loc(S_i) = \bigcup_{p
    \in P_i} \delta_i(p)$, for all $i=0,1,\ldots,n$. Second, let $\ell
  \arrow{s}{} m$ be an edge in $S$. Then $\ell \in dom(h_i)$, for some
  $i=0,1,\ldots,n$, and therefore $m \in Img(h_i)$.  Hence there
  exists $p \in P_i$ such that $\ell,m \in \delta_i(p)$. Third, let $q
  \in P$ be on a path from $p$ to $r$, with $p,r \in P$. We
  distinguish two cases:
  \begin{itemize}
  \item $q=\epsilon$, $p=i.p'$ and $r=j.r'$, with $p' \in P_i$ and $r'
    \in P_j$, $0 \leq i < j \leq n$. Then $\delta(p) \cap \delta(r)
    \subseteq \{\iota(x) ~|~ x \in \vec{z}\} = \delta(q)$
  \item $q = i.q'$, $q' \in P_i$, and $p = i.p'$, $r = i.r'$ for some
    $i=0,1,\ldots,n$. Then $\delta_i(p') \cap \delta_i(r') \subseteq
    \delta_i(q')$ by the fact that $\delta_i$ is a tree decomposition
    of $S_i$, and hence $\delta(p) \cap \delta(r) \subseteq \delta(q)$
  \end{itemize}
  Finally, $\card{\delta(0.p)} \leq \max(\len{\varphi},\card{PVar})$
  for all $p \in P_0$, $\card{\delta(i.p)} \leq \card{P}^{var}$ for
  all $p \in P_i$, $i=1,\ldots,n$, and $\card{\delta(\epsilon)} \leq
  \card{\vec{z}}$. Hence $\card{\delta(p)} \leq
  \max(\card{\vec{z}},\len{\varphi},\card{PVar},\card{\mathcal{P}}^{var})$,
  for all $p \in dom(\delta)$. \qed

\begin{proposition}\label{heap-dom}
  For any state $S = \langle s,h \rangle$ and any interpretations
  $\iota : LVar_{mso} \rightharpoonup_{fin} Loc$ and $\nu : LVar_{mso}
  \rightharpoonup_{fin} 2^{Loc}$ of the first- and second-order
  variables, respectivelly, we have $S,\iota,\nu \models_{mso} Heap(X)
  \iff \nu(X)=dom(h)$.
\end{proposition}
\proof{By definition of $Heap(X)$, $\nu(X)$ is the set of locations
  $\ell \in Loc$ such that $h_s(\ell) \neq \bot$, for some $s =
  1,\ldots,\card{Sel}$. But this is exactly the definition of
  $dom(h)$, by Def. \ref{state}. \qed}


\vspace*{\baselineskip}\noindent The following lemma says that an MSO
formula obtained as a translation of a basic spatial SL formula is
true in a state $S$ if and only if it is true on any extension of $S$.
\begin{lemma}\label{basic-extension}
  Let $\sigma$ be a basic spatial SL formula, $S = \langle s, h
  \rangle$ be any state, $\iota : LVar_{mso} \rightharpoonup_{fin}
  Loc$, $\nu : LVar_{mso} \rightharpoonup_{fin} 2^{Loc}$ be
  interpretations of first and second-order variables,
  respectivelly. Then, for any state $S'$, such that $S \uplus S'$ is
  defined, we have:
  $$S,\iota,\nu[X \leftarrow dom(h)] \models_{mso}
  \overline{\sigma}(X) \iff S \uplus S', \iota, \nu[X \leftarrow
    dom(h)] \models_{mso} \overline{\sigma}(X)$$
\end{lemma}
\proof{By induction on the structure of $\sigma$. \qed}

\vspace*{\baselineskip}\noindent{\bf Proof of Lemma
  \ref{basic-sl-mso}:} By induction on the structure of $\varphi$. The
most interesting case is the separating conjunction, i.e. $\varphi
\equiv \sigma_1 * \sigma_2$ for two spatial SL formulae $\sigma_1$ and
$\sigma_2$.

  \vspace*{\baselineskip}\noindent ``$\Rightarrow$'' $S,\iota
  \models_{sl} \sigma_1 * \sigma_2$ if and only if there exist two
  states $S_1 = \langle s, h_1 \rangle$ and $S_2 = \langle s, h_2
  \rangle$ such that $S_i,\iota \models_{sl} \sigma_i$, for both
  $i=1,2$ and $S = S_1 \uplus S_2$. By the induction hypothesis we
  have that $S_i,\overline{\iota}, \nu[Y_i \leftarrow dom(h_i)]
  \models_{mso} \overline{\sigma_i}(Y_i) \wedge Heap(Y_i)$, and by
  Lemma \ref{basic-extension}, we have $S_1 \uplus S_2,
  \overline{\iota}, \nu[Y_i \leftarrow dom(h_i)] \models_{mso}
  \overline{\sigma_i}(Y_i)$, for both $i=1,2$. Hence:
  \[\begin{array}{lcl}
  S_1 \uplus S_2, \overline{\iota}, 
  \nu[Y_1 \leftarrow dom(h_1)][Y_2 \leftarrow dom(h_2)] 
  & \models_{mso} & \overline{\sigma_1}(Y_1) \wedge \overline{\sigma_2}(Y_2) 
  \\ 
  S_1 \uplus S_2, \overline{\iota}, 
  \nu[X \leftarrow dom(h_1) \cup dom(h_2)] & 
  \models_{mso} & \exists Y_1 \exists Y_2 ~.~ \overline{\sigma_1}(Y_1) 
  \wedge \overline{\sigma_2}(Y_2) \wedge \Pi(Y_1,Y_2,X)
  \\
  S_1 \uplus S_2, \overline{\iota}, 
  \nu[X \leftarrow dom(h_1) \cup dom(h_2)] & 
  \models_{mso} & \overline{\varphi}(X)
  \end{array}\]
  By Proposition \ref{heap-dom}, we obtain further:
  $$S_1 \uplus S_2, \overline{\iota}, \nu[X \leftarrow dom(h_1) \cup
    dom(h_2)] \models_{mso} Heap(X)$$ hence $S, \overline{\iota},
  \nu[X \leftarrow dom(h)] \models_{mso} \overline{\varphi}(X) \wedge
  Heap(X)$.

  \vspace*{\baselineskip}\noindent ``$\Leftarrow$'' If
  $S,\overline{\iota}, \nu[X \leftarrow dom(h)] \models_{mso} \exists
  Y_1 \exists Y_2 ~.~ \overline{\sigma_1}(Y_1) \wedge
  \overline{\sigma_2}(Y_2) \wedge \Pi(Y_1,Y_2,X) \wedge Heap(X)$, then
  there exists two sets of locations, call them $L_1$ and $L_2$, such
  that $L_1 \cap L_2 = \emptyset$ and $L_1 \cup L_2 = dom(h)$, such
  that $S,\overline{\iota},\nu[Y_i \leftarrow L_i] \models_{mso}
  \overline{\sigma_i}(Y_i)$, for both $i=1,2$. Let $h_1$ and $h_2$ be
  the restrictions of $h$ to $L_1$ and $L_2$, respectivelly, and
  define $S_i = \langle s, h_i \rangle$, for both $i=1,2$. Clearly
  $S_1 \uplus S_2 = S$. By Lemma \ref{basic-extension} we have $S_i,
  \overline{\iota},\nu[Y_i \leftarrow dom(h_i)] \models_{mso}
  \overline{\sigma_i}(Y_i)$. By Proposition \ref{heap-dom} we have,
  moreover that $S_i, \overline{\iota},\nu[Y_i \leftarrow dom(h_i)]
  \models_{mso} Heap(Y_i)$, for both $i=1,2$. Applying the induction
  hypothesis, we obtain $S_i,\iota \models_{sl} \sigma_i$, for both
  $i=1,2$, hence $S,\iota \models_{sl} \varphi$. \qed

\vspace*{\baselineskip}\noindent {\bf Proof of Lemma \ref{twa-run}:}
The proof relies on the following claim:
\begin{claim}
For any two positions $p,r \in dom(t)$ such that either (i) $p$ is a
child of $r$, (ii) $r$ is a child of $p$, or (iii) $p=r$ we have:
$$\mbox{$x^p = y^r$ occurs in $\phi_t$} ~\iff~ A_{\mathcal{P}}
~\mbox{moves in one step from $\langle p,q^{var}_x \rangle$ to
$\langle r,q^{var}_y \rangle$}$$
\end{claim}
\proof{We give the proof for the second case, the rest of the cases 
being similar. Assume that $r = p.k$, for some
$k \in \mathcal{D}_+(\mathcal{P})$. ``$\Rightarrow$'' By the
definition of $\phi_t$, $x^{p} = y^{p.k}$ occurs in $\phi^p_t$ only if
$t(p) \equiv R_{ij}$, $(tail(R_{ij}))_k =
P_{i_k}(y_1,\ldots,y_{n_{i_k}})$, where $P_{i_k}(x_{i_k,1}, \ldots,
x_{i_k,n_{i_k}})$ is the corresponding definition in $\mathcal{P}$,
$x \equiv x_{i_k,\ell}$, and $y\equiv y_\ell$, for some
$\ell=1,\ldots,n_{i_k}$. In this case we have
$(q^{var}_{x_{i_k},\ell},k) \in \Delta(q^{var}_{y_\ell},R_{ij}^s,\tau)$,
for all $s\in\mathcal{D}(\mathcal{P})$ and all $\tau \in \Sigma_{\mathcal{P}} \cup \{?\}$, and the conclusion
follows. ``$\Leftarrow$'' By definition, $A_{\mathcal{P}}$ has a
transition rule $(q^{var}_y,k) \in \Delta(q^{var}_x,\sigma^s,\tau)$ for an
$s\in\mathcal{D}(\mathcal{P})$ only
if $\sigma \equiv R_{ij}$, $(tail(R_{ij}))_k \equiv
P_\ell(y_1,\ldots,y_{n_\ell})$, $y\equiv x_{\ell,j}$ and $x \equiv
y_j$, for some $j=1,\ldots,n_{\ell}$. In this case, the equality
$x^p=y^r$ occurs in $\phi^p_t$. \qed}

\noindent''$\Rightarrow$'' $x^p=y^r$ is implied by $\phi_t$ only if 
there exists a path $p = p_1, \ldots, p_n = r$ in $dom(t)$, and
variables $x \equiv z_1, z_2, \ldots, z_{n-1},z_n \equiv y \in
LVar_{sl}$, such that $z_i^{p_i} = z_{i+1}^{p_{i+1}}$ occurs in
$\phi_t$, for all $i=1,\ldots,n-1$. By the above claim,
$A_{\mathcal{P}}$ has a run from $\langle p, q^{var}_x \rangle$ to
$\langle r, q^{var}_t \rangle$ along this path. ``$\Leftarrow$'' If
$A_{\mathcal{P}}$ has a run from $\langle p,q^{var}_x \rangle$ to
$\langle r,q^{var}_y \rangle$ over $t$, there exist a sequence of
positions $p = p_1, p_2, \ldots, p_{n-1}, p_n = q \in dom(t)$ and
variables $x \equiv z_1, z_2, \ldots, z_{n-1}, z_n \equiv y$ such that
$A_{\mathcal{P}}$ moves in one step from $\langle p_i,
q^{var}_{z_i} \rangle$ to $\langle p_{i+1},
q^{var}_{z_{i+1}} \rangle$. By the above claim, there exist equalities
$z_i^{p_i} = z_{i+1}^{p_{i+1}}$ occurring in $\phi_t$, hence $x^p =
y^r$ is a consequence of $\phi_t$. \qed

\vspace*{\baselineskip}\noindent {\bf Proof of Lemma \ref{predicate-sl-mso}:}
``$\Rightarrow$'' If $S,\iota \models_{sl} P_i(x_1,\dots,x_n)$
  then $S,\iota \models_{sl} \phi_t$ for some unfolding tree $t \in
  \mathcal{T}_i(\mathcal{P})$. By induction on the structure of $t$,
  one can build a bijective tree $\mu_0 : dom(t) \rightarrow dom(h)$,
  and define sets $S_{ij}^d = \{\ell \in dom(h) ~|~
  t(\mu_0^{-1}(\ell)) \equiv R_{ij} ~\mbox{and}~ \exists p \in dom(t)
  ~.~ \mu_0^{-1}(\ell) = p.d\}$ for all $d \in
  \mathcal{D}_+(\mathcal{P})$ and $S_{ij}^{-1} = \{\ell \in dom(h) ~|~
  \mu_0^{-1}(\ell) = \epsilon ~\mbox{and}~ t(\epsilon) \equiv
  R_{ij}\}$. Let $\overline{\iota}' = \overline{\iota}[r \leftarrow
    \mu_0(\epsilon)]$ and $\nu : LVar_{mso} \rightharpoonup_{fin}
  2^{Loc}$ be any interpretation of second order variables such that
  $\nu(X_{ij}^d) = S_{ij}^d$ and $\nu(T)=dom(h)$. By Proposition
  \ref{heap-dom}, we have $S,\overline{\iota}',\nu \models_{mso}
  Heap(T)$. Next, we prove the following conditions:
  \begin{enumerate}
  \item $S,\overline{\iota}',\nu \models_{mso}
    backbone_i(r,\seq{X},T)$
  \item $S,\overline{\iota}',\nu \models_{mso} inner\_edges(r,\seq{X},T)$
  \item $S,\overline{\iota}',\nu \models_{mso} no\_double\_alloc(r,\seq{X},T)$
  \item $S,\overline{\iota}',\nu \models_{mso}
    param_{i,j}(r,\seq{X},T)$, for all $j = 1, \ldots, n_i$
  \end{enumerate}

  \noindent (1) Let $\lambda_0 : dom(t) \rightarrow
  \{X_1,\ldots,X_m\}$ be a function defined as $\lambda(p) = X_j$ iff
  $\mu_0(p) \in S_j$. It is immediate that $\nu(X_i) = \{\ell \in dom(h)
  ~|~ \lambda_0(\mu_0^{-1}(\ell)) = X_i\}$. By Lemma \ref{tree-mso}
  (Appendix \ref{app:tree-mso}) it follows that:
  $$S,\overline{\iota}',\nu \models_{mso} tree(r,\seq{X},T) \wedge
  X_i^{-1}(r)$$ To show: $$S,\overline{\iota}',\nu \models_{mso}
  succ\_labels(\seq{X})$$ let $X_{ij}^k$ be an arbitrary variable
  from $\seq{X}$ and let $\ell \in S_{ij}^k$ be a location. Hence
  $t(\mu_0^{-1}(\ell)) \equiv R_{ij}$. Suppose that $tail(R_{ij}) =
  \langle P_{k_1}, \ldots, P_{k_{r_{ij}}} \rangle$, and let $m =
  \mu_0(\mu_0^{-1}(\ell).d)$ for some arbitrary $d \in
  \{1,\ldots,r_{ij}\}$. Clearly $m \in S^d_{k_d j}$, for some $j =
  1,\ldots,n_{k_d}$. One can now easily check
  that: $$S,\overline{\iota}'[x \leftarrow \ell][y \leftarrow m], \nu
  \models_{mso} succ_d(x,y,\seq{X})$$ which concludes this point.

  \noindent (2) If $S,\overline{\iota}',\nu \models_{mso}\exists
  \seq{Y} ~.~ \Phi_{A_{\mathcal{P}}}(r,\seq{X},T,\seq{Y})$ then there
  exist sets $U_1,\ldots,U_k \subseteq Loc$ such that:
  $$S,\overline{\iota}',\nu[\seq{Y} \leftarrow \seq{U}] \models_{mso}
  \Phi_{A_{\mathcal{P}}}(r,\seq{X},T,\seq{Y})$$ By Lemma
  \ref{tree-mso}, there exists a unique prefix-closed set $P \subseteq
  \nat^*$ a unique bijective tree $\mu : P \rightarrow dom(h)$ and a
  unique tree $\lambda : P \rightarrow \{X_1,\ldots,X_m\}$ meeting the
  three properties of Lemma \ref{tree-mso}. Since $dom(t)$, $\mu_0$
  and $\lambda_0$ meet the requirements of Lemma \ref{tree-mso}, it
  turns out that $P=dom(t)$, $\mu=\mu_0$ and $\lambda=\lambda_0$. By
  Lemma \ref{twa-mso}, $A_{\mathcal{P}}$ has an accepting run $\pi$
  over $\lambda_0$, such that $U_j = \nu(Y_j) = \{\mu_0(p) ~|~
  \mbox{$\langle p, q_j \rangle$ occurs on $\pi$}\}$. Let $\ell, m$ be
  two arbitrary locations such that: $$S,\overline{\iota}'[x
    \leftarrow \ell][y \leftarrow m], \nu[\seq{Y} \leftarrow \seq{U}]
  \models_{mso} Y_s^{sel}(x) \wedge Y_f(y)$$ By Lemma \ref{twa-mso},
  there exist positions $p_1, p_2 \in dom(t)$, such that
  $\mu_0(p_1)=\ell$ and $\mu_0(p_2)=m)$, and variables $\alpha, \beta
  \in LVar_{sl}$ such that $A_{\mathcal{P}}$ has a run: $$\langle p_1,
  q_s^{sel} \rangle, \langle p_1, q^{var}_\alpha \rangle \ldots
  \langle p_2, q^{var}_\beta \rangle, \langle p_2, q_f \rangle$$ over
  $\lambda_0$, and implicitly, over $t$. Notice that, by the
  definition of $A_{\mathcal{P}}$, $q_s^{sel}$ and $q_f$ occur exactly
  once on each accepting run, and moreover, $q_f$ is the final state
  on the run.

  Hence, by Lemma \ref{twa-run}, $S,\iota \models_{sl} \alpha^{p_1} =
  \beta^{p_2}$. Since $\ell$ and $m$ are allocated at $p_1$ and $p_2$
  in $t$, respectivelly, there exists an edge $\ell \arrow{s}{} m$ in
  $S$. We have, subsequently: $$S,\overline{\iota}'[x \leftarrow
    \ell][y \leftarrow m],\nu \models_{mso} edges_s(x,y)$$ which
  concludes this point.

  \noindent (3) By contradiction, let us suppose that there exist two
  distinct locations $\ell$ and $m$ such that:
  $$S,\overline{\iota}'[x \leftarrow \ell][y \leftarrow m],\nu[\seq{Y}
    \leftarrow \seq{U}] \models_{mso}
  \Phi_{B_{\mathcal{P}}}(r,\seq{X},T,\seq{Y}) \wedge Y_0(x) \wedge
  Y_f(y)$$ By an argument similar to the one from point (2), there
  exist two variables $\alpha,\beta \in LVar_{sl}$ such that
  $B_{\mathcal{P}}$ has a run:
  $$\langle \mu_0^{-1}(\ell), q_0 \rangle, \langle \mu_0^{-1}(\ell),
  q^{var}_\alpha \rangle, \ldots, \langle \mu_0^{-1}(m), q^{var}_\beta
  \rangle, \langle \mu_0^{-1}(m), q_f \rangle$$ over $\lambda_0$, or
  equivalently, over $t$. By the definition of $B_{\mathcal{P}}$,
  $\alpha^{\mu_0^{-1}(\ell)} = \beta^{\mu_0^{-1}(m)}$ is a consequence
  of $\phi_t$ and moreover both $\alpha$ and $\beta$ are allocated at
  positions $\mu_0^{-1}(\ell)$ and $\mu_0^{-1}(m)$ in $t$,
  respectivelly. The latter facts contradict with the hypothesis that
  $S,\iota \models_{sl} \phi_t$, since, in this case, $\phi_t$ would
  not be satisfiable, according to the semantics of SL.

  \noindent (4) This point follows the case split in the definition of
  $param_{i,j}$ and is proved among the same lines as point (2)
  above. 

  \noindent ``$\Leftarrow$'' If $S,\overline{\iota},\nu[T \leftarrow
    dom(h)] \models_{mso}
  \overline{P_i}(\overline{x_1},\ldots,\overline{x_k},T)$, then there
  exists a location $\ell \in Loc$ and sets $L_1,\ldots,L_m \subseteq
  Loc$ such that:
  \begin{enumerate}[i.]
    \item $S,\overline{\iota}[r \leftarrow \ell],\nu[\seq{X}
      \leftarrow \seq{L}][T \leftarrow dom(h)] \models_{mso}
      backbone_i(r,\seq{X},T)$
    \item $S,\overline{\iota}[r \leftarrow \ell],\nu[\seq{X}
      \leftarrow \seq{L}][T \leftarrow dom(h)] \models_{mso}
      inner\_edges(r,\seq{X},T)$
    \item $S,\overline{\iota}[r \leftarrow \ell],\nu[\seq{X}
      \leftarrow \seq{L}][T \leftarrow dom(h)] \models_{mso}
      no\_double\_alloc(r,\seq{X},T)$
    \item $S,\overline{\iota}[r \leftarrow \ell],\nu[\seq{X}
      \leftarrow \seq{L}][T \leftarrow dom(h)] \models_{mso}
      param_{i,j}(r,\seq{X},T)$, for all $1 \leq i \leq n$ and all $1
      \leq j \leq n_i$
  \end{enumerate}
  For simplicity, we denote $\overline{\iota}' = \iota[r \leftarrow
    \ell]$ and $\nu' = \nu[\seq{X} \leftarrow \seq{L}][T \leftarrow
    dom(h)]$ in the rest of this proof. By (i) and Lemma
  \ref{tree-mso}, there exist a set $P \subseteq \nat^*$, a bijective
  tree $\mu : P \rightarrow dom(h)$, and a tree $\lambda : P
  \rightarrow \seq{X}$ such that $L_i = \{\ell \in dom(h) ~|~
  \lambda(\mu^{-1}(\ell)) = X_i \}$. Since each variable $X_{ij}$ from
  $\seq{X}$ corresponds one-to-one to the rule $R_{ij}$ from
  $\mathcal{P}$, we can build a tree $t : P \rightarrow \seq{R}$ as
  $t(p) \stackrel{def}{=} R_{ij}$ iff $\lambda(p) = X_{ij}$, for all
  $p \in P$. Since, by (i): $$S,\overline{\iota}',\nu' \models_{mso}
  succ\_labels(\seq{X}) \wedge X_i^{-1}(r)$$ we obtain that $t$ is an
  unfolding tree (Def. \ref{unfolding}), and moreover $t \in
  \mathcal{T}_i(\mathcal{P})$. It remains to be shown that $S,\iota
  \models_{sl} \phi_t$. To this end, we extend $\iota$ to an
  assignment $\iota_t : LVar_{sl}^P \rightarrow Loc$ such that:
  \begin{enumerate}
  \item $\iota_t(x^\epsilon) = \iota(x)$, for all $x \in LVar_{sl}$
  \item $\iota_t(x^{p_1}) = \iota(y^{p_2})$, for all $p_1,p_2 \in P$
    and $x,y \in LVar_{sl}$ such that $\phi_t \rightarrow x^{p_1} =
    y^{p_2}$
  \item for all $p \in P$, if $head(t(p)) \equiv x \mapsto (y_1,
    \ldots, y_s)$ then $\iota_t(x^p) \arrow{i}{} \iota_t(y_i^p)$ in
    $S$, for all $i=1,\ldots,s$, and moreover, there are no other
    outgoing edges from $\iota_t(x^p)$ in $S$
  \end{enumerate}
  Given $\overline{\iota}$, $\mu$ and $t$, defined above, we define
  $\iota_t$ as follows:
  \begin{itemize}
  \item $\iota_t(x_j^\epsilon) \stackrel{def}{=}
    \overline{\iota}(\overline{x_j})$, for all $j=1,\ldots,k$
  \item $\iota_t(x^p) \stackrel{def}{=} \mu(p)$, for all $p \in P$
    such that $head(t(p)) \equiv x \mapsto (\ldots)$
  \item for all $x^p \in LVar^P$ not assigned previously,
    $\iota_t(x^p) \stackrel{def}{=} \iota_t(y^q)$ if and only if
    $\phi_t \rightarrow x^p = y^q$ and $y^q$ is assigned by one of the
    above points
  \end{itemize}
  Since $\mathcal{P}$ is established, every existentially
  quantified variable $x^p \in LVar^P$ that occurs in $\phi_t$ is
  connected to an allocated variable $y^q$, i.e. $head(t(q)) \equiv
  y^q \mapsto (\ldots)$, by a path of equalities $x^p = z_1^{p_1} =
  \ldots = z_n^{p_n} = y^q$ all occurring in $\phi_t$. Hence $\iota_t$
  assigns locations to all existentially quantified variables in
  $\phi_t$. Clearly, $\iota_t$ satisfies points (1) and (2) above. To
  show that $\iota_t$ meets point (3), fix an arbitrary position $p_0
  \in P$ such that $head(t(p_0)) \equiv x \mapsto (y_1, \ldots,
  y_s)$. Observe first that, by (i), $\iota_t(x^{p_0})$ has no
  outgoing edges $\iota_t(x^{p_0}) \arrow{i}{} \ell$, for any $i >
  s$. For the rest, let us fix some arbitrary $1 \leq i_0 \leq s$ and
  show that $\iota_t(x^{p_0}) \arrow{i_0}{} \iota_t(y^{p_0}_{i_0})$ is
  an edge in $S$. There are two cases:
  \begin{itemize}
  \item $y_k$ is an existentially quantified variable of $\phi_t$
  \item $y_k$ is a parameter $x_j$ of the predicate $P_i$
  \end{itemize}
  We shall carry out the proof only in the first case, the reasoning
  being similar in the second. As previously discussed, if $y_k$ is
  existentially quantified, there exists a sequence of equalities
  $y_{i_0}^{p_0} = z_1^{p_1} = \ldots = z_n^{p_n}$ occurring in $\phi_t$,
  such that $z_n$ is allocated by $head(t(p_n))$. By Lemma
  \ref{twa-run}, $A_{\mathcal{P}}$ has a run from $\langle p_0,
  q^{var}_{y_{i_0}} \rangle$ to $\langle p_n, q^{var}_{z_n} \rangle$ over
  $t$. Hence $A_{\mathcal{P}}$ has also a run: 
  $$\langle p_0, q^{sel}_s \rangle, \langle p_0, q^{var}_{y_k}
  \rangle, \ldots, \langle p_n, q^{var}_{z_n} \rangle, \langle p_n,
  q_f \rangle$$ over $t$, and, equivalently, a loop-free run $\pi$
  over $\lambda$. Let $U_j = \{\mu(p) ~|~ ~\mbox{$\langle p, q_j
    \rangle$ occurs on $\pi$}\}$. By Lemma \ref{twa-mso}, we obtain:
  $$S,\overline{\iota}',\nu'[\seq{Y} \leftarrow \seq{U}] \models_{mso}
  \Phi_{A_{\mathcal{P}}}(r,\seq{X},T,\seq{Y})$$ and moreover,
  $\mu(p_0) \in U^{sel}_s$ and $\mu(p_n) \in U_f$, where $U^{sel}_s$
  and $U_f$ are the sets of locations corresponding to the states
  $q^{sel}_s$ and $q_f$, respectivelly. We obtain, further: 
  $$S,\overline{\iota}'[x \leftarrow \mu(p_0)][y \leftarrow
    \mu(p_n)],\nu' \models_{mso} \exists \seq{Y} ~.~
  \Phi_{A_{\mathcal{P}}}(r,\seq{X},T,\seq{Y}) \wedge Y_s^{sel}(x)
  \wedge Y_f(y)$$ By (ii), we obtain: $$S,\overline{\iota}'[x
    \leftarrow \mu(p_0)][y \leftarrow \mu(p_n)],\nu' \models_{mso}
  edge_s(x,y)$$ hence the conclusion follows. \qed

\begin{lemma}\label{predicate-extension}
  Let $P_i(x_{i,1},\ldots,x_{i,n})$ be a predicate of a recursive
  definition system $\mathcal{P}$, $S = \langle s, h \rangle$ be a
  state, and $\iota : LVar_{mso} \rightharpoonup_{fin} Loc$ and $\nu :
  LVar_{mso} \rightharpoonup_{fin} 2^{Loc}$ be interpretations of
  first and second-order variables, respectivelly. Then, for any state
  $S'$, such that $S \uplus S'$ is defined, we have: $$S,\iota,\nu[T
    \leftarrow dom(h)] \models_{mso}
  \overline{P_i}(\overline{x_{i,1}},\ldots,\overline{x_{i,n}},T)$$ $$\iff S \uplus
  S',\iota,\nu[T \leftarrow dom(h)] \models_{mso}
  \overline{P_i}(\overline{x_{i,1}},\ldots,\overline{x_{i,n}},T)$$
\end{lemma}
\proof{The proof is done by inspection of
  $\overline{P}(\overline{x_1},\ldots,\overline{x_n},T)$. Namely we
  need to prove the following equivalences, for some $\ell \in Loc$
  and sets $S_1,\ldots,S_m$ corresponding to the variables
  $X_1,\ldots,X_m$:
  \[\begin{array}{rclcrcl}
  S,\iota',\nu' & \models_{mso} & backbone_i(r,\seq{X},T) & \iff &
  S \uplus S',\iota',\nu' & \models_{mso} & backbone_i(r,\seq{X},T) \\
  S,\iota',\nu' & \models_{mso} & inner\_edges(r,\seq{X},T) & \iff &
  S \uplus S',\iota',\nu' & \models_{mso} & inner\_edges(r,\seq{X},T) \\
  S,\iota',\nu' & \models_{mso} & no\_double\_alloc(r,\seq{X},T) & \iff &
  S \uplus S',\iota',\nu' & \models_{mso} & no\_double\_alloc(r,\seq{X},T) \\
  S,\iota',\nu' & \models_{mso} & param_{ij}(r,\seq{X},T) & \iff &
  S \uplus S',\iota',\nu' & \models_{mso} & param_{ij}(r,\seq{X},T) \\
  \end{array}\]
  for all $1 \leq j \leq n$, where $\iota'=\iota[r \leftarrow \ell]$
  and $\nu'=\nu[\seq{X}\leftarrow\seq{S}][T \leftarrow dom(h)]$. These
  equivalences can be proved by case analysis. \qed}

\vspace*{\baselineskip}\noindent{\bf Proof of Theorem \ref{slrd-mso}:}
Let us first consider the case $k=0$, i.e. $\varphi$ is a basic SL
formula $\varphi \equiv \exists \vec{z} ~.~ \phi(\vec{y}_0)$. By Lemma
\ref{basic-sl-mso}, for any state $S=\langle s,h \rangle$, we have
that $S \models_{sl} \varphi$ if and only if $S,\iota \models_{mso}
\overline{\varphi}(X) \wedge Heap(X)$, where $\iota$ is any
interpretation such that $\iota(X)=dom(h)$. Hence $S \models_{mso}
\exists X ~.~ \overline{\varphi}(X) \wedge Heap(X)$. Dually, if $S
\models_{mso} \exists X ~.~ \overline{\varphi}(X) \wedge Heap(X)$,
then $S,\iota[X \leftarrow L] \models_{mso} \overline{\varphi(X)}
\wedge Heap(X)$, where $L \subseteq Loc$ is a set of locations. By
Proposition \ref{heap-dom}, we have $L=dom(h)$. Hence $S,\iota
\models_{sl} \varphi$, by Lemma \ref{basic-sl-mso}.

  \vspace*{\baselineskip}\noindent The case $k > 0$ is dealt with by
  induction on $k$. For $k=1$ we have $\varphi \equiv \exists \vec{z}
  ~.~ \phi(\vec{y}_0) * P(\vec{y}_1)$. ``$\Rightarrow$'' If $S
  \models_{sl} \exists \vec{z} ~.~ \phi(\vec{y}_0) * P(\vec{y}_1)$
  then $S,\iota \models_{sl} \phi(\vec{y}_0) * P(\vec{y}_1)$ for some
  interpretation $\iota : LVar_{sl} \rightharpoonup_{fin} Loc$, such
  that $\iota(x) \neq \bot$, for all $x \in \vec{z}$. Hence there
  exists two states $S_1 = \langle s, h_1 \rangle$ and $S_2 = \langle
  s, h_2 \rangle$ such that $S_1 \uplus S_2 = S$, and moreover $S_1,
  \iota \models_{sl} \phi(\vec{y}_0)$ and $S_2, \iota \models_{sl}
  P(\vec{y}_1)$. Applying Lemma \ref{basic-sl-mso} and
  \ref{predicate-sl-mso}, respectivelly, we obtain:
  \[\begin{array}{lcl}
  S_1, \overline{\iota}[X_0 \leftarrow dom(h_1)] & \models_{mso} & 
  \overline{\phi}(\overline{\vec{y}_0}, X_0) \wedge Heap(X_0)
  \\
  S_2, \overline{\iota}[X_1 \leftarrow dom(h_2)] & \models_{mso} &
  \overline{P}(\overline{\vec{y}_1}, X_1) \wedge Heap(X_1)
  \end{array}\]
  where $\overline{\iota} : LVar_{mso} \rightharpoonup_{fin} (Loc \cup
  2^{Loc})$ is an interpretation meeting the requiremenets of both
  Lemma \ref{basic-sl-mso} and \ref{predicate-sl-mso}. Applying Lemma
  \ref{basic-extension} and \ref{predicate-extension}, respectivelly,
  we obtain:
  \[\begin{array}{lcl}
  S, \overline{\iota}[X_0 \leftarrow dom(h_1)] & \models_{mso} & 
  \overline{\phi}(\overline{\vec{y}_0}, X_0)
  \\
   S, \overline{\iota}[X_1 \leftarrow dom(h_2)] & \models_{mso} &
   \overline{P}(\overline{\vec{y}_1}, X_1)
  \end{array}\]
  Since $S_1 \uplus S_2 = S$, we obtain:
  \[\begin{array}{lcl}
  S, \overline{\iota}[X \leftarrow dom(h)] & \models_{mso} & \exists X_0
  \exists X_1 ~.~ \overline{\phi}(\overline{\vec{y}_0}, X_0) \wedge
  \overline{P}(\overline{\vec{y}_1}, X_1) \wedge \Pi(X_0,X_1,X) 
  \\
  S, \overline{\iota}[X \leftarrow dom(h)] & \models_{mso} & 
  \exists \vec{z} \exists X_0 \exists X_1 ~.~ \overline{\phi}(\overline{\vec{y}_0}, X_0) 
  \wedge \overline{P}(\overline{\vec{y}_1}, X_1) \wedge \Pi(X_0,X_1,X)
  \end{array}\]
  and by Proposition \ref{heap-dom} we also have $S,
  \overline{\iota}[X \leftarrow dom(h)] \models_{mso} Heap(X)$. The
  conclusion follows. ``$\Leftarrow$'' If $S \models_{mso} \exists X
  ~.~ \overline{\varphi}(X) \wedge Heap(X)$, then for any
  interpretation $\overline{\iota} : LVar_{mso} \rightharpoonup_{fin}
  (Loc \cup 2^{Loc})$ we have $S,\overline{\iota}[X \leftarrow dom(h)]
  \models_{mso} \overline{\varphi}(X)$, by Proposition
  \ref{heap-dom}. Hence there exists $L_1, L_2 \subseteq dom(h)$, such
  that $L_1 \cap L_2 = \emptyset$, $L_1 \cup L_2 = dom(h)$, and:
  \[\begin{array}{lcl}
  S,\overline{\iota}[X_0 \leftarrow L_1] & \models_{mso} & \overline{\phi}(\overline{\vec{y}_0},X_0) \\
  S,\overline{\iota}[X_1 \leftarrow L_2] & \models_{mso} & \overline{P}(\overline{\vec{y}_1},X_1)
  \end{array}\]
  Let $h_1, h_2$ be the restrictions of $h$ to $L_1$, $L_2$,
  respectivelly, and $S_1 = \langle s, h_1 \rangle$, $S_2 = \langle s,
  h_2 \rangle$. Clearly $S = S_1 \uplus S_2$. By Lemma
  \ref{basic-extension} and \ref{predicate-extension}, respectivelly,
  we have that:
  \[\begin{array}{lcl}
  S_1,\overline{\iota}[X_0 \leftarrow L_1] & \models_{mso} & \overline{\phi}(\overline{\vec{y}_0},X_0) \\
  S_2,\overline{\iota}[X_1 \leftarrow L_2] & \models_{mso} & \overline{P}(\overline{\vec{y}_1},X_1)
  \end{array}\]
  and by Lemma \ref{basic-sl-mso} and \ref{predicate-sl-mso}, respectivelly, we obtain:
  \[\begin{array}{lcl}
  S_1,\iota & \models_{sl} & \phi(\vec{y}_0) \\
  S_2,\iota & \models_{sl} & P(\vec{y}_1)
  \end{array}\]
  for an intepretation $\iota : LVar_{sl} \rightharpoonup_{fin} Loc$
  meeting the conditions of Lemma \ref{basic-sl-mso} and
  \ref{predicate-sl-mso}. Hence $S,\iota \models_{sl} \phi(\vec{y}_0)
  * P(\vec{y}_1)$, which leads to $S \models_{sl} \exists \vec{z} ~.~
  \phi(\vec{y}_0) * P(\vec{y}_1)$. 

  \vspace*{\baselineskip}\noindent The induction step follows a
  similar argument. \qed

\end{document}